\documentclass[fleqn,usenatbib]{mnras}

\usepackage{newtxtext,newtxmath}
\DeclareRobustCommand{\VAN}[3]{#2}
\let\VANthebibliography\thebibliography
\def\thebibliography{\DeclareRobustCommand{\VAN}[3]{##3}\VANthebibliography}

\usepackage{amsmath}
\usepackage[T1]{fontenc}
\usepackage{graphicx}
\usepackage{enumitem}
\usepackage{xcolor}
\usepackage{hyperref}
\usepackage{gensymb}
\usepackage{CJK}
\usepackage{tabularx}
\usepackage{orcidlink}
\usepackage{soul} 
\hypersetup{
    unicode, 
    colorlinks=true,
    linkcolor=linkcolor,
    citecolor=linkcolor,
    filecolor=linkcolor,
    urlcolor=linkcolor,
}
\definecolor{linkcolor}{rgb}{0.0,0.3,0.5}

\defcitealias{Ness2025}{Paper~I}

\title[ACACIAS II: Sculptor abundances]{ACACIAS II: Element abundances for 39 stars in the dwarf galaxy Sculptor}

\author[Cornejo-Cárdenas et al.]{Anell Cornejo-Cárdenas,$^{1}$\thanks{E-mail: anell.cornejocardenas@anu.edu.au}
Sven Buder,$^{1}$\thanks{Australian Research Council DECRA Fellow}\orcidlink{0000-0002-4031-8553}
Melissa K. Ness,$^{1}$\orcidlink{0000-0001-5082-6693}
and J. Trevor Mendel$^{1}$\orcidlink{0000-0002-6327-9147}
\\
$^{1}$Research School of Astronomy and Astrophysics, Australian National University, Canberra, ACT 2611, Australia
}

\date{Accepted 2026 September 3. Received 2026 August 13; in original form 2026 May 7}

\pubyear{2026}

\begin{document}
\label{firstpage}
\pagerange{\pageref{firstpage}--\pageref{lastpage}}
\maketitle

\begin{abstract}
Element abundances in nearby galaxies provide insights into their chemical enrichment histories and place the Milky Way and its neighbours in a broader cosmological context. These systems can be studied through individual stars, whose resolved spectra enable the construction of a chemical abundance census of the Local Group. We present the second stage of the ACACIAS program, which aims to build the first homogeneous dataset of stellar abundances across dwarf galaxies spanning a range of masses for comparison with Milky Way properties. We analyse low-resolution (R = 1800), low-SNR ($\sim 25$) spectra obtained with the Multi Unit Spectroscopic Explorer (MUSE) for 39 stars in the Sculptor dwarf spheroidal galaxy. Using the data-driven method $The~Cannon$, we combine high-fidelity labels from the APOGEE survey with MUSE-like LAMOST spectra to analyse the MUSE data. This enables the determination of [Mg/Fe], [Mn/Fe], [C/Fe], [N/Fe], and [Al/Fe] on the same abundance scale as APOGEE, with precisions of $\sigma = 0.09-0.37~\mathrm{dex}$. Comparing different model configurations, we show that [Mg/Fe], [Mn/Fe], and [C/Fe] are best recovered from targeted spectral windows, whereas stellar parameters and [Al/Fe] benefit from full-spectrum information with telluric masking. This reflects the complementary strengths of restricted-window models, which recover abundances from direct spectral features, and full-spectrum models, which exploit correlations with other labels when diagnostic features are weak or absent. Our approach demonstrates the potential of low-resolution, low-SNR stellar spectra to build more homogeneous datasets for extragalactic archaeology to better understand the chemical enrichment diversity of the Local Group.
\end{abstract}


\section{Introduction}
\label{sec:introduction}

Through the study of stellar motions and chemical compositions, Galactic archaeology provides a window into the formation history of our own Galaxy \citep{Freeman2002,Helmi2008,Bland-Hawthorn2016}. In the $\Lambda$CDM cosmological framework, massive galaxies like the Milky Way are assembled through the continuous accretion and merging of smaller systems, whose stellar populations become dispersed throughout the Galaxy, leaving traces that we can study today \citep{White1978,Bullock2005}. The stellar populations from these accreted dwarf galaxies thus constitute the fundamental building blocks of the Milky Way \citep{Tolstoy2009,Helmi2020}.

Observations from the \textit{Gaia} satellite \citep{GaiaCollaboration2016, GaiaCollaboration2018} complemented by ground-based spectroscopic surveys such as APOGEE \citep{Majewski2017}, GALAH \citep{DeSilva2015}, RAVE \citep{Steinmetz2006}, and LAMOST \citep{Cui2012}, among others, have revealed the chemical diversity of the Milky Way's stellar halo. These observations expose the numerous accretion events that have shaped, and continue to shape, its structure. For example, there is compelling evidence that the Sagittarius dwarf galaxy is currently being tidally disrupted by the Milky Way \citep{Ibata1994, Majewski2003}. In addition, the Milky Way is thought to have experienced a major merger event approximately 8–10 Gyr ago, commonly referred to as \textit{Gaia}-Sausage-Enceladus (GSE), which disrupted the proto-disk and contributed significantly to the present-day stellar halo \citep{Belokurov2018, Helmi2018, Haywood2018, Mackereth2020, Naidu2020}.

While such disrupted systems provide information about past accretion events, intact surviving satellites offer complementary insights into both the future and past evolution of the Milky Way. On one hand, these dwarf galaxies represent potential future building blocks, as they may eventually be accreted and dissolved into the Milky Way's halo, as is currently happening with the Sagittarius dwarf galaxy \citep{Helmi2020}. On the other hand, they serve as present-day analogues of the progenitors that contributed to past merger events \citep{Tolstoy2009,Helmi2018}. Therefore, characterizing their chemical abundance patterns provides templates that can be used to identify stars from disrupted progenitors now scattered throughout the Milky Way's halo.

\citet{Naidu2022} conducted a census of the logarithmic element abundance ratios of iron to hydrogen and $\alpha$-process elements such as magnesium relative to iron, [Fe/H] and [$\alpha$/Fe], for both surviving and disrupted dwarf galaxies around the Milky Way, finding that these systems span $-2.5 < [\rm Fe/H] < -0.5$ and $-0.2 < [\alpha/\rm Fe] < 0.4$ dex. This census reveals the diversity in metallicity and $\alpha$-enhancement among dwarf galaxies, reflecting their different star formation efficiencies and chemical enrichment histories. For the surviving satellites, the analysis is largely based on metallicity measurements from the literature compilation of \citet{Kirby2013}, derived from resolved medium-resolution (Keck/DEIMOS, with $R\approx 6500$) spectroscopy of individual red giant stars. However, this study is limited to 13 surviving galaxies, despite more than 50 known dwarf galaxies orbiting the Milky Way \citep{Munoz2018}. Expanding chemical characterization to these additional satellites could yield key information into the assembly history of the Milky Way and its accreted building blocks.

Recent advances in instrumentation have enabled resolved spectroscopy of individual stars in nearby galaxies. High-resolution observations with instruments like VLT/UVES and VLT/GIRAFFE have targeted smaller samples of bright stars in the nearest systems \citep[e.g.][]{Hill2019, Skuladottir2019}. Additionally, integral field spectrographs such as MUSE \citep{Bacon2010} have enabled spatially resolved spectroscopy without the need for pre-selection of individual targets, allowing the simultaneous observation of large numbers of stars. Large observational efforts such as the oMEGACat \citep{Nitschai2023} project and crowded‑field 3D spectroscopy campaigns with MUSE (e.g., \citealt{Roth2018}) are now exploiting data-cubes to build extensive catalogues of resolved stellar spectra in dense stellar environments in the Milky Way and other nearby galaxies.

In the near future, next-generation adaptive optics–assisted spectrographs, such as the integral-field instrument HARMONI \citep{Thatte2021} and the multi-object spectrograph HRMOS \citep{Magrini2023}, are expected to extend these capabilities to much higher spatial resolution and sensitivity, enabling detailed spectroscopy of individual stars in more distant and more crowded galaxies. In parallel, adaptive-optics assisted visible instruments such as MAVIS \citep{Rigaut2020,Rigaut2021} will enable the study of stellar populations and chemical abundances in more distant and higher-redshift galaxies.

This wealth of integral‑field spectroscopic data motivates the development of robust automated methods to derive stellar parameters and chemical abundances. In particular, data-driven approaches have been developed to learn empirical relationships between spectra and stellar labels (such as stellar parameters or abundances) from training sets with known parameters.  When trained on well-characterised reference stars, data-driven models such as \textit{The Cannon} \citep{Ness2015} allow the data themselves to define empirical label–flux correlations across all spectral pixels. Information not explicitly represented in line-by-line physical models can therefore contribute to the inference. This leads to more precise label estimates, less dependence on uncertain atomic data, and a substantially lower computational cost than full spectrum synthesis (for a review see Ness 2026, in press). However, a careful vetting process is required, including validation of the learned correlations and potentially the application of regularisation or masking strategies to suppress unphysical correlations \citep{Casey2016, Buder2018}. These methods have enabled the efficient and homogenous analysis of large spectroscopic datasets. By exploiting their similarity in spectral resolution and wavelength coverage, machine‑learning models have even been used to infer stellar labels from LAMOST and MUSE spectra by \citet{Wang2022} and \citet[][hereafter Paper~I]{Ness2025}.

These studies and instruments have confirmed the diversity among Local Group dwarf galaxies in mass, morphology, and star formation histories. However, comparing chemical abundances across different galaxies and with the Milky Way requires homogeneous data collection and uniform analysis methods in order to minimize systematic biases. The "extra-galactic ArChAeology with Cannon Inferred Abundances from resolved integral field unit Spectra" \citepalias[ACACIAS;][]{Ness2025} program addresses this challenge by applying a data-driven method to low-quality spectra ($R \sim 1800$, SNR $> 10$) of stars in the Milky Way and other galaxies, enabling consistent and direct comparisons of their chemical enrichment histories. In \citetalias{Ness2025}, we use \textit{The Cannon} to recover element abundances of red giant branch (RGB) stars in the dwarf irregular galaxy NGC~6822 with a stellar mass of $10^8\,\mathrm{M_\odot}$ \citep{McConnachie2012} and mean metallicity of [Fe/H] $\sim -1.0$ dex.

Building on this concept, the present work extends the ACACIAS framework to the Sculptor dwarf spheroidal galaxy, a lower-mass \citep[$2.3 \times 10^6\,\mathrm{M_\odot}$][]{McConnachie2012} and lower-metallicity system with [Fe/H] $\sim -1.6$ dex ($\sim$3\% solar). Our study also addresses a key limitation of data-driven models: the lack of transparency regarding which spectral features drive abundance predictions. By incorporating physically motivated wavelength selections and exploring multiple model configurations (Section~\ref{subsec: cannon-training-details}), we provide a more interpretable approach that enables robust chemical tagging and demonstrates the applicability of \textit{The Cannon} to intermediate-resolution extragalactic datasets. By establishing a consistent framework for measuring detailed chemical abundances from low-resolution spectra, this work contributes to a more complete and uniform census of the Milky Way's building blocks and their distinct chemical signatures across both disrupted and surviving dwarf galaxies, a crucial step toward reconstructing the Galaxy's hierarchical assembly history.

This paper is organized as follows. Section~\ref{sec:data} describes the observational data, including MUSE spectra of Sculptor stars and the construction of our training set. Section~\ref{sec: cannon-methodology} presents \textit{The Cannon} methodology. In Section~\ref{sec: analysis}, we present the our model optimization strategies and the results for Sculptor stellar abundances. In Section~\ref{sec:discussion} we discuss our results and compare the abundances of Sculptor with those from the literature as well as those of NGC~6822 from \citetalias{Ness2025}. Finally, in Section~\ref{sec:conclusions} we summarize the work.

\section{Data} \label{sec:data}

\begin{figure*}
    \centering
    \includegraphics[width=18cm]{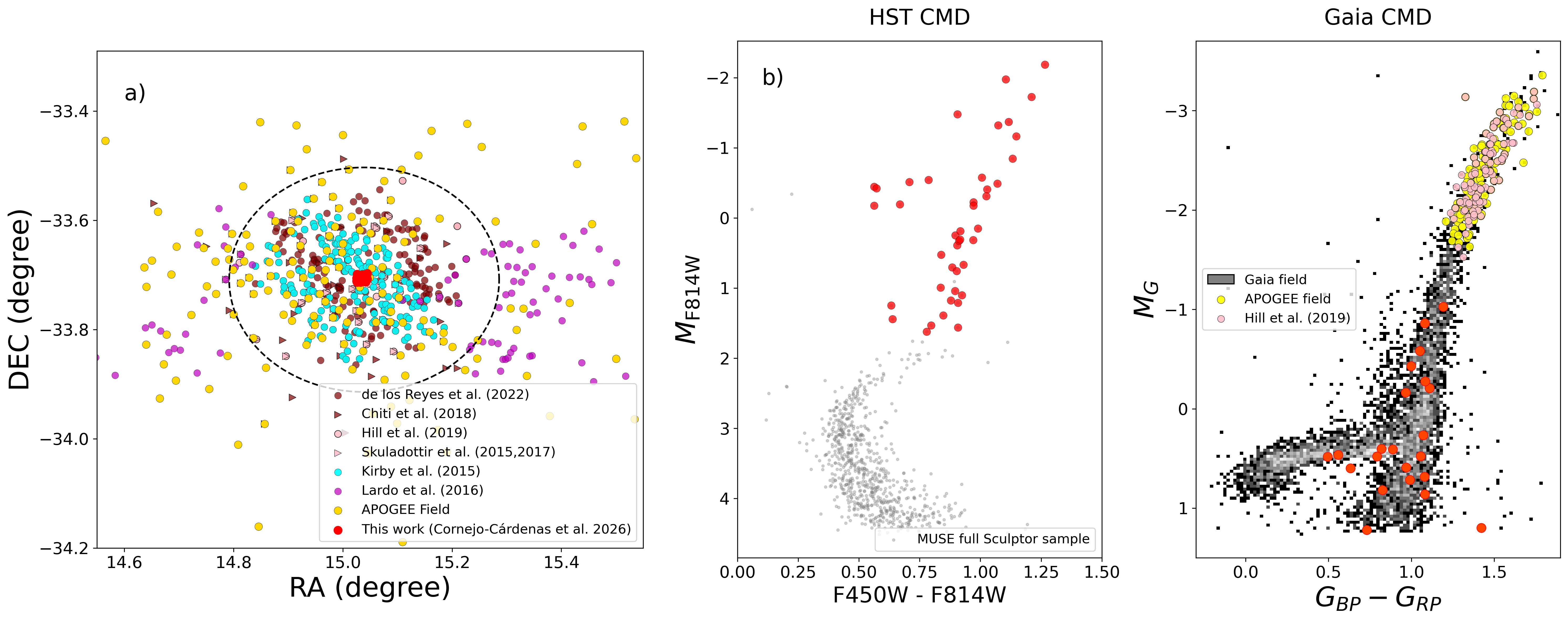}
    \caption{(a) Spatial distribution of stars: selected Sculptor stars (this work, red) and literature samples \citep{delosReyes2022,Chiti2018, Hill2019, Skuladottir2015,Skuladottir2017,Kirby2015,Lardo2016}. APOGEE targets are shown as yellow circles.  (b) HST color–magnitude diagram, showing absolute $F814W$ magnitude versus $F450W-F814W$ color. Gray points show the full MUSE catalog, while red circles indicate the selected sample analysed in this work. (c) Gaia color–magnitude diagram showing absolute $G$ magnitude versus $G_{\rm BP}-G_{\rm RP}$ color. The gray background shows the Gaia field in the Sculptor region, while pink and yellow circles indicate stars from \citet{Hill2019} and APOGEE targets, as in panel (a). Red circles show the stars from our selected sample with a Gaia source identified through a cross-match using a matching radius of $2\arcsec$.}
    \label{fig:RaDec_CMD}
\end{figure*}

\subsection{Observations of Sculptor}

Sculptor is a dwarf spheroidal (dSph) galaxy and a satellite of the MW, with a stellar mass of $\approx 2.3 \cdot10^6 ~ \rm M_\odot$ (\citealt{McConnachie2012}, and references therein). It lies at a heliocentric distance of $83.9 \pm 1.5$ kpc \citep{Martin-Vazquez2015} and was the first Local Group dwarf galaxy in which a metallicity gradient from more metal-rich in the centre to more metal-poor in the outskirts was identified \citep{Tolstoy2004}. Its star formation history is dominated by an old ($>10$ Gyr), metal-poor stellar population \citep{deBoer2012}, with a mean metallicity of [Fe/H] = -1.6 \citep{Tolstoy2001}.

This dwarf galaxy has been extensively studied through spectroscopic observations spanning a wide range of spectral resolutions, from high-resolution analyses of individual RGB stars \citep{Hill2019, Skuladottir2015, Skuladottir2017} to medium- and low-resolution surveys \citep{Kirby2015, Lardo2016, Chiti2018, delosReyes2022}. Together, these studies provide a comprehensive view of the physical and chemical properties of Sculptor’s stellar populations and a valuable comparison for our lower resolution and signal-to-noise analysis with MUSE, albeit of different stars in the less central regions of Sculptor (see Fig.~\ref{fig:RaDec_CMD}a).

\subsubsection{MUSE spectra and HST photometry}\label{subsec: museHST_details}

The central regions of Sculptor (see Fig.~\ref{fig:RaDec_CMD}a) were observed using MUSE in the WFM-AO configuration as part of ESO Programmes 0101.D-300(A), 0102.D-0372(A), and 105.208V.001 (PI: Brinchmann) for a total of 9x900s exposures. Raw MUSE data and their associated calibrations were download from the ESO archive and processed using v2.10.14 of the ESO MUSE instrument pipeline \citep{Weilbacher2020} and \textsc{pymusepipe}\footnote{https://github.com/emsellem/pymusepipe} following standard procedures. Given the relatively high number of bright, unresolved (stellar) sources in each exposure, we align the individual frames to a common reference using a simple cross-correlation of white-light images in x and y. We additionally tested for potential rotation offsets between frames, however all frames appeared to be aligned to <0.04 deg compared to their nominal rotator positions, and so no further corrections were applied. The final combined cube was constructed using a simple mean of in individual reconstructed exposures.

Individual stellar spectra were extracted using PampelMUSE \citep{Kamann2013,Kamann2018} via reference photometry from the Hubble Space Telscope following the procedures outlined in \citetalias{Ness2025}. Our photometric catalogues consist of HST/WFPC2 imaging in F450W and F814W from GO-10888 (PI: Cole); we use the F814W imaging as our reference image for \textit{PampelMUSE}. We extract a total of 816 individual stellar spectra extending down to $m_{\rm F814W}$ = 24.2 [Vega]. Of those 816, 81 have SNR > 10. The MUSE spectra are calibrated on a vacuum wavelength scale and were convolved to match the LAMOST spectral resolution assuming the wavelength-dependent MUSE resolution derived by \cite{Bacon2017}.

From this MUSE sample of Sculptor stars (see Fig.~\ref{fig:RaDec_CMD}b), we identified 42 RGB stars using the color-magnitude diagram (CMD) constructed from HST photometry in the F450W and F814W filters. Specifically, these stars have colors of $0.4 < (\text{F450W} - \text{F814W}) < 2.0$ and apparent magnitudes of F814W $< 21.3$. The red points in Fig. \ref{fig:RaDec_CMD}b indicate the Sculptor stars selected for this study.

Additionally, for completeness, we cross-matched our selected sample with Gaia DR3 using a matching radius of $2\arcsec$. The red circles in Fig.~\ref{fig:RaDec_CMD}c show the Gaia sources corresponding to the stars in our selected sample. The gray background represents the distribution of Gaia sources in the Sculptor field. The HST and Gaia photometry shown in Fig. \ref{fig:RaDec_CMD} has not been corrected for extinction.

\subsubsection{Literature data of Sculptor spectra}

Fundamental stellar parameters, including $T_{\mathrm{eff}}$, $\log g$, and metallicity [Fe/H], as well as $\alpha$-element abundances such as [Mg/Fe], have been derived across both high- and low-resolution datasets \citep{Skuladottir2015, Hill2019, Kirby2015, Lardo2016}. In contrast, some elemental abundances, including [C/Fe] \citep{Kirby2015, Lardo2016, Chiti2018}, [N/Fe] \citep{Lardo2016}, and [Mn/Fe] \citep{delosReyes2022}, have so far been reported primarily from low-resolution studies, highlighting the potential of low-resolution spectra to measure elements that may be challenging or time-intensive to determine at high resolution.

Literature data for this comparison were compiled from the VizieR catalog service \citep{Ochsenbein2000} by querying the term "Sculptor" and filtering results to ensure they targeted the Sculptor dwarf spheroidal galaxy. For comparison, we additionally considered APOGEE DR17 data \citep{Abdurrouf2022}, obtained by filtering the FIELD column for "Sculptor" entries and applying the following selection criteria: $\log g < 4$ (to be consistent with focus on RGB stars) and, following \cite{Kirby2015}, eliminating stars more than three standard deviations away from the mean radial velocity. Figure \ref{fig:RaDec_CMD}a shows the sky coverage of RGB stars from both the literature studies and APOGEE (yellow points). The black dashed circle marks the effective radius of Sculptor, $r_{\rm eff}=308~\rm pc$ (which encloses half the total luminosity) from \cite{Munoz2018}.

Using the right ascension and declination of our MUSE sample and the \texttt{search\_around\_sky} module from the \texttt{Astropy} package \citep{AstropyCollaboration2013}, we verified that there is no positional overlap between our data and these literature samples. Our observations are located closer to the centre of the Sculptor galaxy (RA = 15\fdg03917, Dec = -33\fdg70889; adopted from the SIMBAD astronomical database, \citealt{Wenger2000}). In contrast, previous studies have targeted different regions: for instance, \citet{Lardo2016} focused on the outer parts of the galaxy, while \citet{Chiti2018} covered intermediate regions. Other works, such as \citet{Kirby2015}, \citet{delosReyes2022}, \citet{Skuladottir2015, Skuladottir2017}, and \citet{Hill2019}, include stars in the inner regions but do not overlap with our MUSE sources.

\subsection{Training set input: APOGEE labels \& LAMOST spectra}

Before measuring stellar abundances in the Sculptor dwarf galaxy, we must first train \textit{The Cannon} model on how to connect spectrum flux with stellar parameters and abundances. For this purpose, and following similar procedures
outlined in \cite{Ho2017,Ho2017b} and \citetalias{Ness2025}, we construct a training set consisting of RGB stars observed by both the APOGEE and LAMOST surveys, leveraging APOGEE's high-quality stellar labels and LAMOST's optical spectra that overlap in wavelength coverage and spectral resolution with our MUSE observations of Sculptor. However, these previous studies primarily focused on the most luminous red giant branch stars ($\log g < 2.5$) when defining their training samples. In order to achieve a more representative coverage of the stellar populations observed in Sculptor, we therefore construct an expanded training set that also includes less luminous stars, extending into the subgiant region ($\log g < 4$).

\subsubsection{APOGEE}\label{subsec: APOGEE}
Following the approach of the \citetalias{Ness2025}, we adopt stellar labels from the Apache Point Observatory Galactic Evolution Experiment (APOGEE) Data Release 17 \citep{Abdurrouf2022}. APOGEE is a high-resolution ($R \sim 22500$) near-infrared spectroscopic survey covering the wavelength range $\lambda \sim 1.51$--$1.70~\mu$m (H-band), designed to conduct a homogeneous chemical census of Milky Way stellar populations \citep{Majewski2017}.  The APOGEE Stellar Parameters and Chemical Abundances Pipeline (ASPCAP; \citealt{GarcíaPérez2016}) derives stellar atmospheric parameters ($T_{\rm eff}$, $\log g$, microturbulence velocity $v_{\rm mic}$) and individual element abundances through a $\chi^2$ minimization technique that compares observed spectra to synthetic spectral libraries.
APOGEE has also obtained spectra of individual stars in Milky Way satellite dwarf galaxies \citep{Hasselquist2021}, deriving chemical abundances from high signal-to-noise ratio observations (SNR $> 70$) using the ASPCAP pipeline. 

To ensure high-quality labels of APOGEE data, we follow the same selection quality criteria as \citetalias{Ness2025}. We select stars with \texttt{ASPCAPFLAG} = 0 (no critical ASPCAP warnings) and exclude stars with large relative uncertainties by requiring $\sigma(T_{\rm eff})/T_{\rm eff} < 0.05$ and $\sigma(\log g)/\log g < 0.05$. Additionally, following \citetalias{Ness2025} criteria, we restrict the sample to stars with low extinction, $0 < A_K < 0.4$ mag, to avoid heavily reddened sight-lines and implement the following abundances requirements: 
\begin{flalign*}
& -3 < \mathrm{[Fe/H]} < 0.5 & \\
& -3 < \mathrm{[C/Fe]}, ~ \mathrm{[Mn/Fe]}, ~\mathrm{[Al/Fe]} < 2 & \\
& -3 < \mathrm{[N/Fe]}, ~\mathrm{[Mg/Fe]} < 3 & \\
& ~~~\mathrm{[Fe/H]_\text{err}} ~, ~\mathrm{[Mn/Fe]_\text{err}} < 0.1 & \\
& ~~~\mathrm{[Mg/Fe]_\text{err}} ~, ~\mathrm{[Al/Fe]_\text{err}} < 0.2 & \\
& ~~~\mathrm{[C/Fe]_\text{err}} ~, ~\mathrm{[N/Fe]_\text{err}} < 0.2 & 
\end{flalign*}

Compared to \citetalias{Ness2025},  our training set covers a larger extent of the red giant branch, across $\log g < 4.0$ and $3000 < T_{\rm eff} < 5500$ K. This larger range is adopted to provide a good coverage of the full stellar parameter range of the Sculptor stars.

These selection criteria result in a parent sample of $N = 331,633$ APOGEE RGB stars with high-quality stellar parameters and chemical abundances suitable for training.

\subsubsection{LAMOST} \label{subsec: LAMOST}

The Large Sky Area Multi-Object Fiber Spectroscopic Telescope (LAMOST) is a 4-meter telescope conducting spectroscopic surveys of millions of stars across the Milky Way \citep{Cui2012}. LAMOST low-resolution spectra cover the optical wavelength range $\lambda \sim 3700$--$9000$ Å with a spectral resolution of $R \sim 1800$, with wavelengths reported in vacuum. This wavelength range and spectral resolution make LAMOST spectra comparable to our MUSE observations of Sculptor. Due to this similarity, we use LAMOST DR9 spectra to train \textit{The Cannon} for estimating chemical abundances of RGB stars in Sculptor from their MUSE spectra.

\subsubsection{Cross-match between APOGEE and LAMOST}\label{subsec: APOGEExLAMOST}
To construct our training set, we cross-matched the quality-selected APOGEE DR17 sample with the LAMOST DR9 catalogue using the \texttt{crossmatch\_angular} function from \texttt{astropy} \citep{AstropyCollaboration2013} with a matching radius of $1''$. This cross-match yields $N = 67,216$ stars with both high-quality APOGEE labels and LAMOST optical spectra. Note that we do not apply SNR cuts to the LAMOST spectra at this stage, as the optimal SNR threshold will be determined through the training set optimization presented in Section~\ref{subsec: Best training set selection}.

\begin{figure*}
    \centering
    \includegraphics[width=0.9\linewidth]{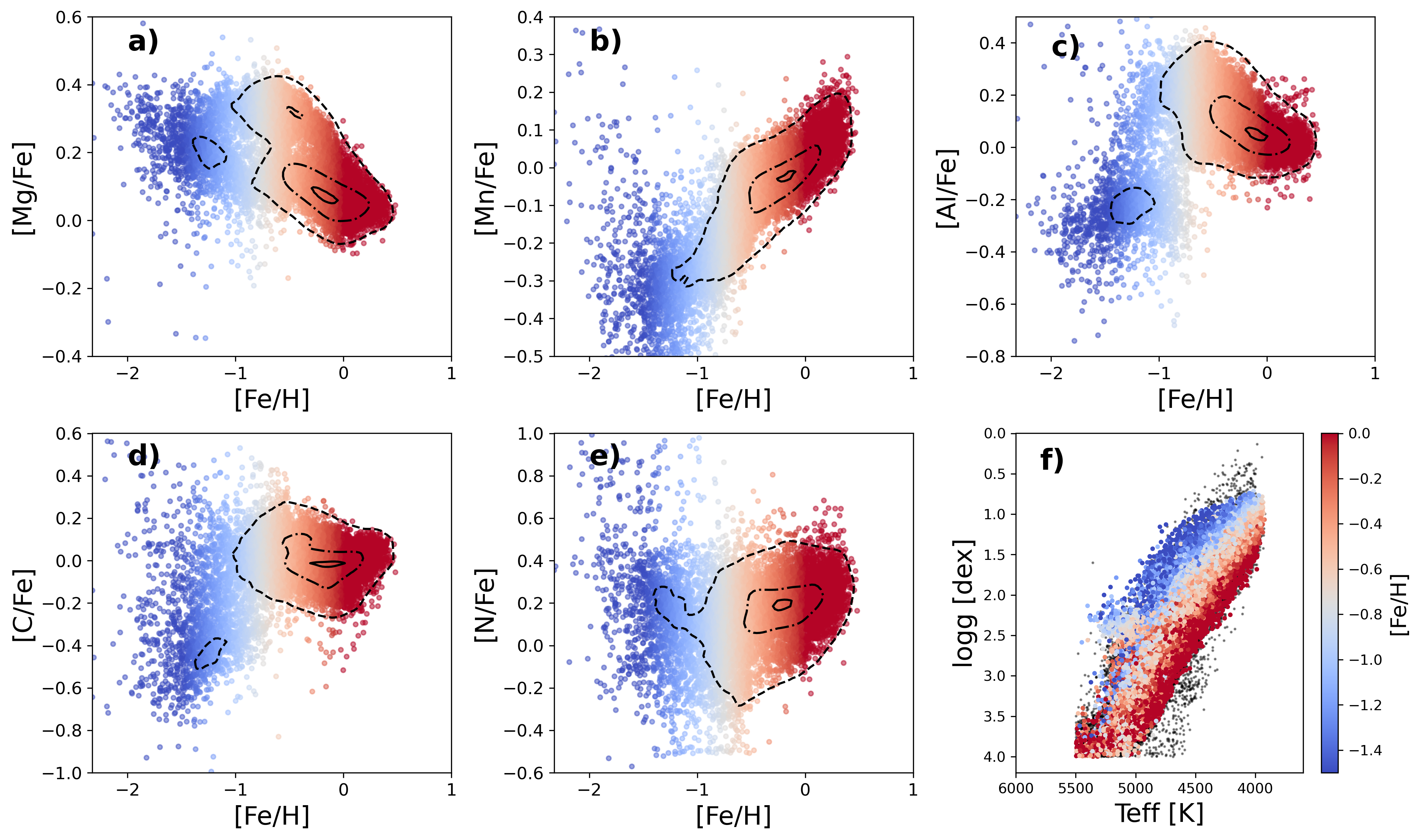}
    \caption{Panels a)-e) show the chemical abundance distributions of giant stars of Mg, Mn, Al, C and N, respectively. Black contours show the density distribution of the full giant sample, enclosing regions that contain 5\%, 50\%, and 95\% of the total probability (dashed, dash-dotted, and solid lines, respectively). The scatter plot shows the final training set (see Section~\ref{subsec: Best training set selection} for details), color-coded by [Fe/H]. Panel f) shows the CMD of the training set color-coded by [Fe/H] }
    \label{fig: trainingset_X/feh}
\end{figure*}

The black contours in Figure~\ref{fig: trainingset_X/feh} show the distribution of this RGB star sample in the $[\rm Fe/H]$--$[X/\rm Fe]$ planes (with $X=$ Mg, Mn, Al, C, and N). The dashed, dash-dotted, and solid lines represent regions containing 5\%, 50\%, and 95\% of the probability density, respectively. The colored points show the distribution of the final training set after applying SNR cuts (described in detail in Section~\ref{subsec: Best training set selection}).

\subsubsection{LAMOST spectra}\label{subsec:data_spectra}

To prepare the LAMOST spectra for use with \textit{The Cannon}, we preprocess them following the requirements outlined by \citet{Ness2015}. This includes shifting the spectra to the rest frame, resampling them onto a common wavelength grid, continuum-normalizing the flux, and propagating the corresponding inverse variances (IVAR), defined as ${\rm IVAR} = 1/\sigma_{\rm flux}^2$.

First, each spectrum is shifted to the rest frame using $\lambda_{\rm rest} = \lambda_{\rm obs}/(1+z)$
where the redshift $z$ is taken from the LAMOST pipeline header. Next, following \citetalias{Ness2025}, the spectra are resampled onto a common wavelength grid of 3559 pixels spanning $\lambda \sim 3900-8800\,\text{\AA}$, with a pixel spacing ranging from $\sim 0.9\,\text{\AA}\mathrm{\,px^{-1}}$ at the blue end to $\sim2.0\,\text{\AA}\mathrm{\,px^{-1}}$ at the red end. The flux and its uncertainty are then continuum-normalized by dividing by a smoothed version of each spectrum, obtained using a Gaussian filter \citep[\textsc{scipy.ndimage.gaussian\_filter1d};][]{Scipy} with a kernel size of $50$ px. Finally, to mitigate the impact of outlier pixels in the high-SNR training spectra, we restrict the normalized flux to the range $0.2 < f_{\rm norm} < 1.3$. Pixels outside this range are assigned the median flux of the spectrum and an uncertainty of $\sigma_{\rm flux} = 999.0$, effectively masking them during training with \textit{The Cannon}.

It is important to note that MUSE spectra of Sculptor stars do not cover the full wavelength range of the LAMOST spectra used for training. While LAMOST extends to shorter wavelengths in the blue, MUSE spectra begin at approximately 4800~\AA. To ensure consistency between training and application, we restrict the model to wavelength regions covered by both datasets (4800~\AA $<\lambda<$ 9000~\AA). This wavelength restriction is straightforward to implement either before the training or afterwards, since the coefficients are solved independently for each wavelength pixel.

\section{Methodology: The Cannon} \label{sec: cannon-methodology}

\textit{The Cannon} is a data-driven method for determining stellar labels, such as effective temperature ($T_{\mathrm{eff}}$); surface gravity ($\log g$); and individual element abundances ([X/Fe]), directly from stellar spectra. This flexible approach has proven effective in transferring information from high-resolution, high SNR surveys (e.g., APOGEE) to lower-resolution, medium SNR surveys (e.g., LAMOST and MUSE). For example, \citet{Ho2017} and \citetalias{Ness2025} showed that this algorithm can recover stellar labels consistent with APOGEE measurements from spectra at lower resolutions, such as those from LAMOST. In this work, we apply \textit{The Cannon} to measure chemical abundances (that are comparable to APOGEE estimations) from MUSE spectra of giant stars in the dwarf spheroidal galaxy Sculptor.

\subsection{Training and testing steps}\label{subsec: cannon-training-details}

In the training step, \textit{The Cannon} takes as input a set of reference objects (in our case, stars) with known labels and their corresponding doppler shift corrected normalized fluxes, and establishes a mapping between them. The goal of this process is for the model to learn how the flux at each wavelength pixel depends on the stellar labels. This training relies on a few key assumptions: (1) stars with identical labels exhibit the same per-pixel flux, (2) the flux at each pixel is independent of that at neighbouring pixels, and (3) the normalized flux at a given pixel can be represented as a polynomial function of the stellar labels (see equation \ref{eq: cannon_f_lambda}).

Following the approach of \citet{Ho2017} and \citetalias{Ness2025}, the flux $f_n$ of a star $n$ at wavelength pixel $\lambda$ can be modelled as a quadratic function of its stellar labels as:
\begin{equation}\label{eq: cannon_f_lambda}
f_{n\lambda} = \Theta_\lambda^T \cdot \ell_n + \text{noise}
\end{equation}

where $\Theta_\lambda$ is a vector of coefficients specific to each pixel, and $\ell_n$ denotes the set of stellar labels. In the training phase, we adopted stellar labels from the APOGEE DR17 survey, consisting of stellar parameters ($T_\text{eff}$, $\log g$, $v_\text{mic}$, and [Fe/H]), elemental abundances ([Mg/Fe], [Mn/Fe], [Al/Fe], [C/Fe], and [N/Fe]), as well as the interstellar extinction $A_{Ks}$. The latter was included as it has been shown to improve the recovery of stellar labels in the optical (e.g., \citealt{Ho2017}; \citealt{Buder2018}). \textit{The Cannon} then models the noise for each star at each wavelength as:

\begin{equation}\label{eq: noise}
    \text{noise}^2 = \sigma_{n,\lambda}^2 + s_\lambda^2
\end{equation}

where $\sigma_{n,\lambda}^2$ is the observational flux uncertainty (varying with wavelength and star), and $s_\lambda$ is the intrinsic scatter of the model at each wavelength, estimated during training. $\sigma_{n,\lambda}^2$ is the observational uncertainty, $s_\lambda$ quantifies how well the model reproduces the observed flux given those uncertainties.

In the training phase, for a given set of fluxes and stellar labels, \textit{The Cannon} determines the coefficients $\Theta_\lambda$ and the intrinsic scatters $s_\lambda^2$ at each pixel by maximizing the single-pixel log-likelihood function (see equation 4 in \citealt{Ness2015} for details). Once trained, the data-driven model can be applied to infer stellar labels for a different test set of observed spectra. In our case, we trained the model using stellar labels from APOGEE and normalized fluxes from LAMOST, and subsequently applied it to resolved MUSE spectra of Sculptor stars to infer their labels.

\textit{The Cannon} allows for masking specific wavelength regions in the spectra, preventing selected labels from contributing to the flux at those pixels by setting all coefficients that are combined with this label at a pixel to zero. This functionality is particularly useful for assessing whether the inferred labels are driven by spectral features related to the label or other correlating features.  For example, it allows us to test whether [Mg/Fe] predictions originate primarily from magnesium absorption lines or from other wavelength regions, such as those of other $\alpha$-process elements that behave similarly to Mg (i.e., Si or Ti). By comparing full-spectrum and masked models, we can distinguish between correlation and causality in this data-driven method.

\subsection{Validation process}\label{subsec: cannon-censoring-validation}

Once trained, a common validation approach is the take-one-star-out test, where the model is iteratively trained on all but one of the $N_\mathrm{ref}$ reference objects and then applied to the excluded object. This procedure is repeated for all stars in the sample, providing independent label estimates for each. However, for large training sets this method is computationally expensive. As an alternative, in this work we adopt the take-25\%-out (or four-fold cross validation) test: the reference set is divided randomly into four groups, each group is excluded in turn while training on the remaining three, and the trained model is then tested on the left-out group. At the end of this process, every reference object has a new set of labels inferred from a model in which it was not included. The robustness of this approach is confirmed by insignificant differences in label predictions between these four cross-validation models and a model trained on the full reference set. The final model used to estimate Sculptor stellar labels is constructed using the complete reference sample.

For estimating precision uncertainties, we adopt a Monte Carlo approach similar to that described in \citetalias{Ness2025}. For each observed spectrum, we generate ten perturbed realizations by modifying the flux at each wavelength according to its reported uncertainty. Specifically, at every pixel we sample a random value from a Gaussian distribution centred on the observed flux and with a standard deviation equal to the corresponding flux error. Each realization therefore represents a perturbation of the observed spectrum consistent with the measured flux uncertainties. The trained model can then be applied to all realizations, yielding a distribution of inferred labels. The standard deviation of this distribution provides an estimate of the precision of the model predictions by propagating the reported pixel-level flux uncertainties through the label-inference procedure.

\section{Analysis} \label{sec: analysis}

\subsection{Training set selection} 
\label{subsec: Best training set selection}

As described in Section~\ref{subsec: APOGEExLAMOST}, we train \textit{The Cannon} using RGB stars observed by both LAMOST and APOGEE. Careful selection of the training set from this sample of giant stars is necessary to achieve accurate abundance predictions, particularly for metal-poor stars similar to those in dwarf galaxies like Sculptor.

\citetalias{Ness2025} constructed their LAMOST-APOGEE training set by selecting giant stars with ${\rm SNR}_{\rm LAMOST} > 100$, or ${\rm SNR}_{\rm LAMOST} > 50$ for more evolved giants with $\log g < 1$. While this approach ensures high-quality spectra for the bulk of the sample, it excludes a significant fraction of stars with $[\rm Fe/H] < -0.5$, which tend to have lower SNR in LAMOST spectra. In particular, the training set used in \citetalias{Ness2025} has a median metallicity of [Fe/H]=-0.29, with the 25th and 75th percentiles at -0.48 and -0.08 dex, respectively. Only 22\% of the sample has [Fe/H]<-0.5.

Disrupted and surviving dwarf galaxies are predominantly metal-poor, with typical metallicities $[\rm Fe/H] \lesssim -0.5$ \citep{Naidu2022}. To construct a training set more representative of the metal-poor stellar populations we aim to study in Sculptor and future-proof for additional applications to other low-mass galaxies, we adopt metallicity-dependent SNR cuts and implement a balancing scheme as described below.

\subsubsection{Metallicity-dependent SNR cuts}
To retain a larger fraction of metal-poor giants while maintaining spectral quality, we apply the following SNR thresholds as a function of metallicity:

\begin{itemize}[leftmargin=2em,labelwidth=2em]
    \item For stars  with $[\rm Fe/H] > -0.5$ : ${\rm SNR}_{\rm LAMOST} > 200$
    \item For stars with $[\rm Fe/H] \leq -0.5$ : ${\rm SNR}_{\rm LAMOST} > 25$
\end{itemize}

After applying these metallicity-dependent SNR cuts, the fraction of stars with [Fe/H]< -0.5 increases to 46\%, and the median metallicity shifts to [Fe/H] = -0.43 with the 25th and 75th percentiles at -0.62 and -0.11 dex, respectively.

\subsubsection{Balancing the training set}

\begin{figure}
    \centering
    \includegraphics[width=1\linewidth]{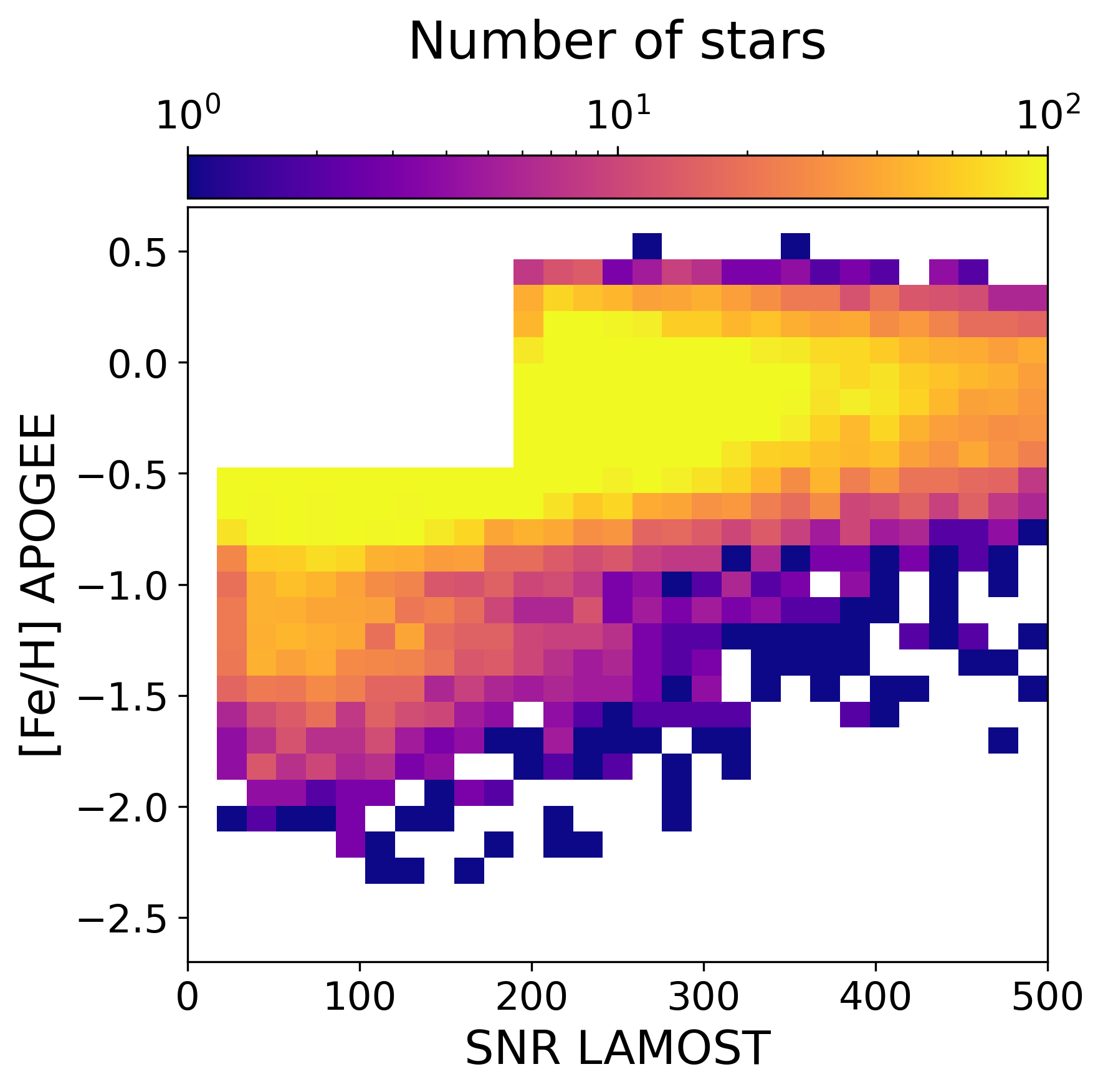}
    \caption{Distribution in SNR (from LAMOST) versus [Fe/H] (from APOGEE) of the selected stars after the metallicity-dependent SNR cuts and balancing process. The color scale indicates the number of stars per bin.} \label{fig: SNR_feh_trainingset}
\end{figure}

\begin{figure*}
    \centering
\includegraphics[width=1\linewidth]{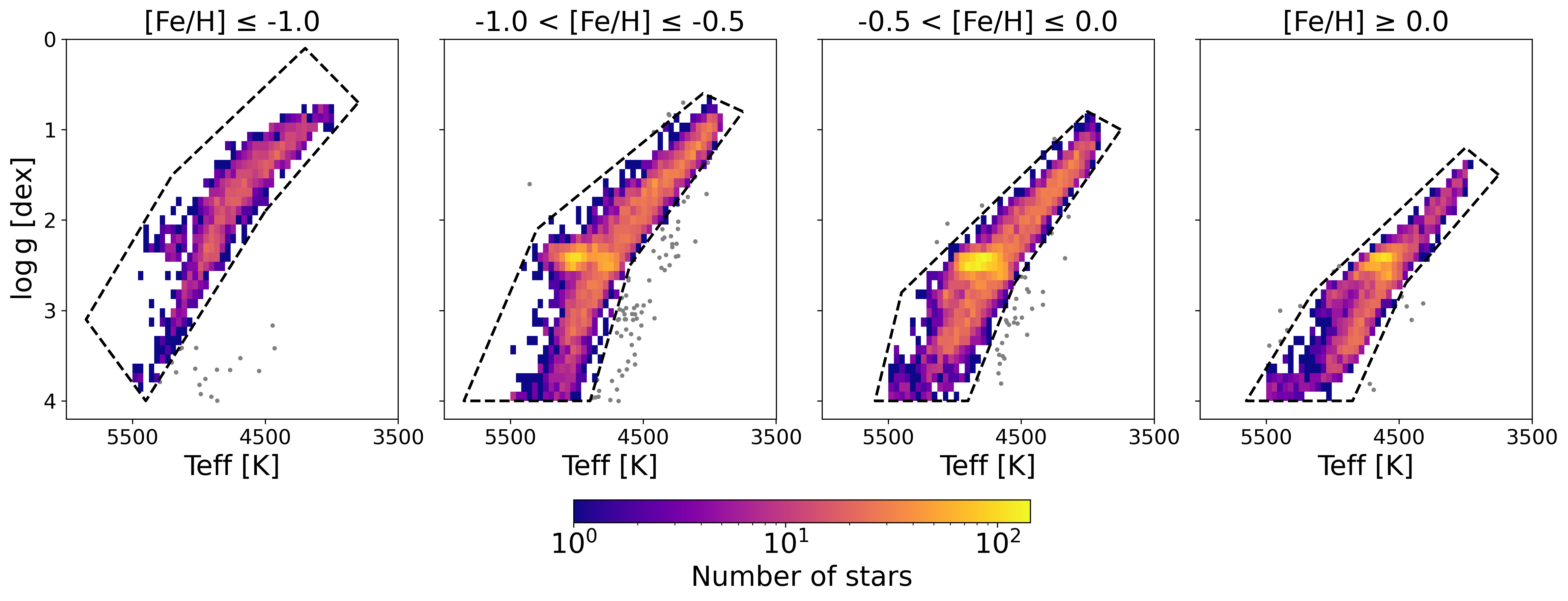}
    \caption{Metallicity-dependent RGB selection in the $T_{\rm eff}$--$\log g$ plane. Each of these panels shows the stellar distribution for one of four metallicity bins, with color indicating the number of stars per bin. Black dashed contours mark the manually defined RGB selection regions used to exclude outliers and non-RGB contaminants (gray dots).}
    \label{fig: [Fe/H]bins-logg-teff}
\end{figure*}

Even after applying metallicity-dependent SNR cuts and extinction limits, the resulting sample remains imbalanced in the ${\rm SNR}$--$[\rm Fe/H]$ space. The more metal-rich bins ([Fe/H]>-0.7) can contain more than a thousand stars, while metal-poor bins are sparsely populated. Such imbalance can bias \textit{The Cannon}'s training via the maximum likelihood estimation and lead to a too strong focus on fitting metal-rich spectra rather than the metal-poor spectra that we are most interested in.

To mitigate this, we implement a balanced sampling strategy. We divide the ${\rm SNR}$--$[\rm Fe/H]$ parameter space into a $30 \times 30$ grid, with 30 bins spanning $0 < {\rm SNR} < 500$ and 30 bins spanning $-2.7 < [\rm Fe/H] < 0.7$ dex as shown in Fig.~\ref{fig: SNR_feh_trainingset}. Within each bin:
\begin{itemize}[leftmargin=2em,labelwidth=2em]
    \item If $N_{\rm bin} > 100$ : randomly select 100 stars
    \item If $N_{\rm bin} \leq 100$ : retain all stars
\end{itemize}

This ensures that sparsely populated metal-poor regions, which are the dominant stellar populations in dwarf galaxies such as Sculptor, receive a more appropriate weight during training, while preventing metal-rich bins from dominating the model optimization. The random selection for populous bins maintains statistical representativeness across all other labels while controlling for sample size imbalance.

This ensures that sparsely populated metal-poor regions, which are the dominant stellar populations in dwarf galaxies such as Sculptor, receive a more appropriate weight during training. At the same time it prevents metal-rich bins from dominating the model training. The random selection for populous bins maintains statistical representativeness across all other labels while controlling for sample size imbalance.

\subsubsection{RGB selection in the $T_{\rm eff}$--$\log g$ plane}

To further refine the training set and remove outliers or non-RGB contaminants that were not picked up by the automatic quality selection of APOGEE, we apply a metallicity-dependent selection in the $T_{\rm eff}$--$\log g$ plane(see Figure~\ref{fig: [Fe/H]bins-logg-teff}). We divide the balanced sample into four metallicity bins ($[\rm Fe/H] \leq -1.0$, $-1.0 < [\rm Fe/H] \leq -0.5$, $-0.5 < [\rm Fe/H] \leq 0.0$, and $[\rm Fe/H] > 0.0$ dex) and manually define RGB selection contours for each bin based on the stellar distribution in the $T_{\rm eff}$--$\log g$ plane. This accounts for the metallicity-dependent shift of the RGB locus and ensures we retain only bona fide RGB stars while excluding outliers and stars in other evolutionary phases. This selection removes $140$ stars from the balanced sample, yielding our final training set of $15,410$ RGB stars.

\begin{figure}
    \centering
\includegraphics[width=1\linewidth]{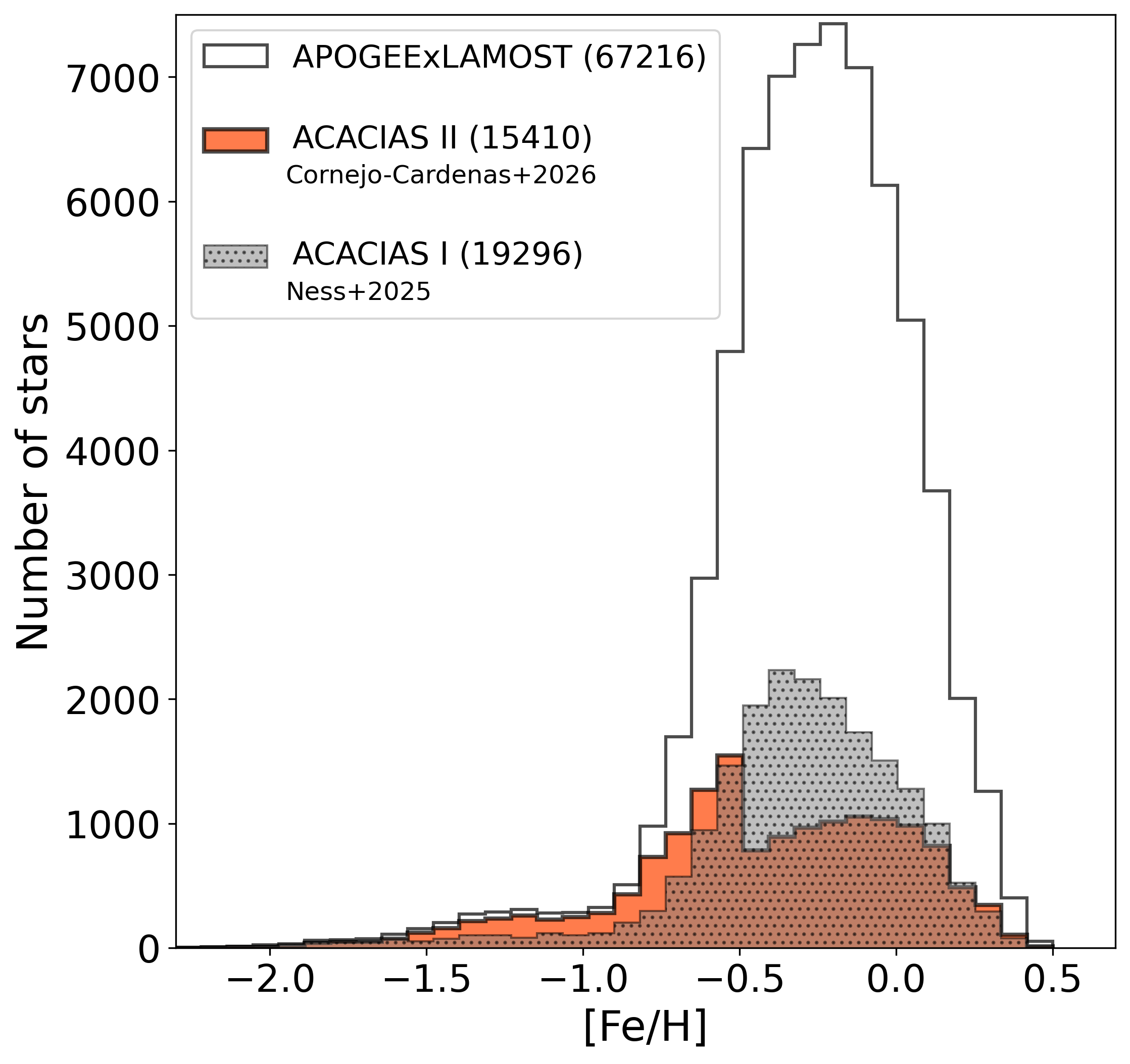}
    \caption{Metallicity distributions of RGB stars. Unfilled histogram: all RGB stars in the APOGEE-LAMOST cross-match (Section~\ref{subsec: APOGEExLAMOST}). Orange histogram: optimized training set used in this work (ACACIAS~II) after applying metallicity-dependent SNR cuts, balanced sampling, and RGB selection (Section~\ref{subsec: final_trainingset_selection}). Gray histogram with black dots: training set from \citetalias{Ness2025}. Numbers in parentheses indicate sample sizes.}
    \label{fig: feh_comparison}
\end{figure}

\subsubsection{Selected Training Set Summary}\label{subsec: final_trainingset_selection}

After applying (1) metallicity-dependent SNR cuts, (2) balanced sampling in the SNR--[Fe/H] space, and (3) metallicity-dependent RGB selection, our final training set consists of $N = 15,410$ RGB stars spanning $-2.3 < [\rm Fe/H] < 0.5$ dex and $25 < {\rm SNR}_{\rm LAMOST} < 500$. Figure~\ref{fig: feh_comparison} compares the metallicity distributions of three samples: the initial APOGEE-LAMOST cross-match (unfilled histogram, $N = 67,216$), the training set from \citetalias{Ness2025} (gray with dot pattern, $N = 19,296$), and our optimized training set (orange, $N = 15,410$). The figure reveals that our training set contains a larger fraction of metal-poor stars compared to \citetalias{Ness2025}. Specifically, 11\% of our training set has [Fe/H] $< -1.0$ dex, compared to only 5\% in \citetalias{Ness2025}. This enhanced metal-poor coverage provides a more representative training sample for analyzing Sculptor. Figure~\ref{fig: trainingset_X/feh} shows the distribution of our training set across multiple abundance planes ([Fe/H] vs. [X/Fe] for X = Mg, Mn, Al, C, N) and in the $T_{\rm eff}$--$\log g$ plane (all color-coded by [Fe/H]), demonstrating good coverage of the parameter space relevant for Sculptor RGB stars.

\subsection{The Cannon Model Configurations}\label{subsec:optimizing the cannon}

Understanding how \textit{The Cannon} predicts stellar labels from input spectra can help improve the accuracy of its estimates. In this work, we trained four configurations of \textit{The Cannon} with different constraints on the training spectra in order to evaluate the algorithm’s sensitivity to specific spectral information. It is important to note that all four models were trained using the same set of 15,410 RGB stars; the only differences arise from the spectral regions that were masked or included during training. These configurations include a Full-spectrum model, an Absorption model, a Telluric model, and a Combined model. Below we describe the specifications of each configuration.

\subsubsection{Full-spectrum model}\label{subsec: full-spectrum model}

This configuration follows the same setup as described in \citetalias{Ness2025}, where all coefficients are fitted during the training process without applying any wavelength masking (see upper panel in Fig.~\ref{fig: theta_2models}). In other words, this model uses the entire spectrum to determine the value of each stellar label and allow us to study correlation.

\subsubsection{Absorption model}\label{subsec: absorption model}

In traditional spectroscopic abundance analysis, element abundances are derived by measuring the equivalent widths or synthesizing the profiles of absorption lines specific to each element. For data-driven methods like \textit{The Cannon}, however, it is not immediately evident whether abundance predictions arise from the physically relevant spectral features or from indirect correlations elsewhere in the spectrum. To investigate this, we construct the Absorption model where we apply element-specific masks that retain only coefficients that involve a specific element, if the wavelength of the pixel is corresponding to known strong absorption lines of said element. For example, when predicting [Mg/Fe], the model uses only spectral regions containing prominent Mg lines (e.g., Mg~I triplet at $\sim$5170 Å), setting all other coefficients to zero. This masking is applied individually for Mg, Mn, Al, C, and N. Iron is excluded from this masking due to its numerous absorption features distributed throughout the optical wavelength range.

Additionally, we applied further modifications to the spectral masks used for estimating Mg, C, and N. Specifically, for magnesium and carbon, in addition to their atomic absorption lines, we also included strong molecular features—such as MgH for Mg, and CH, CN, and C$_2$ for C. For nitrogen, we used spectral regions that include both nitrogen and carbon features. This choice was motivated by the molecular equilibrium between C and N, which are often linked through molecules such as CN \citep{Ting2018}. As a result, the spectral signatures of C and N are difficult to disentangle, since changes in the abundance of one element can affect the molecular line strengths used to infer the other. By including carbon features when estimating nitrogen, \textit{The Cannon} can better capture these molecular dependencies, potentially improving the accuracy of the inferred N abundances.

To construct the wavelength masks we identified the central wavelengths ($\lambda_0$) of the strong absorption lines of each element using the Arcturus line list presented in \cite{Hinkle2000}. The wavelengths in this line list are provided in air and were converted to the vacuum scale to match our MUSE and LAMOST spectra. To define the line width, we considered two main broadening effects: the spectral resolution of the LAMOST spectrograph ($V_\mathrm{LAMOST} = c / \text{R}_\mathrm{LAMOST} = c / 1800$, with $c$ the speed of light) and a rotational velocity component of $V_\mathrm{rot} = 20\,\mathrm{km\,s^{-1}}$ based on the fastest rotators in the sample, chosen to ensure a sufficiently wide coverage around each feature. The final line widths were then computed as:

\begin{equation}\label{eq: delta_lambda}
    \Delta \lambda = \frac{\lambda_0}{c} \cdot \Delta V = \frac{\lambda_0}{c} \cdot \sqrt{(V_\mathrm{LAMOST})^2 + (V_\mathrm{rot})^2}.
\end{equation}

\subsubsection{Telluric model}\label{subsec: telluric model}
Telluric absorption features could introduce systematic biases in the label estimates derived from stellar spectra  \citep[e.g.,][]{Buder2018,Kamann2018b}. To mitigate their impact, we down-weighted the spectral regions affected by significant telluric contamination at 5889.9-5942.6, 6276.6-6315.0, 6867.1-6956.4, 7167.9-7333.7, 7593.6-7703.7, and 8130.4-$8357.0\,\text{Å}$ in the model (see hatched areas in the lower panel of Fig.~\ref{fig: theta_2models}). These wavelength intervals were taken from the line list of \cite{Hinkle2000}, and correspond to the regions with the strongest telluric contamination. To down-weight these spectral regions, we artificially increase the flux uncertainties within those wavelength intervals, thereby reducing their influence during the training process. 

\subsubsection{Combined model}\label{subsec: combined model}

Finally, this model combines the spectral constraints from the Absorption model and the Telluric model. Thus, to estimate the stellar labels, this model considers the strong absorption features of the elements (and molecules for some of them, such as Mg, C, and N; as described in Section~\ref{subsec: absorption model}), and additionally down-weights the influence of the spectral regions affected by telluric absorption (see Section~\ref{subsec: telluric model}).

\begin{figure*}
    \centering
    \includegraphics[width=1.02\linewidth]{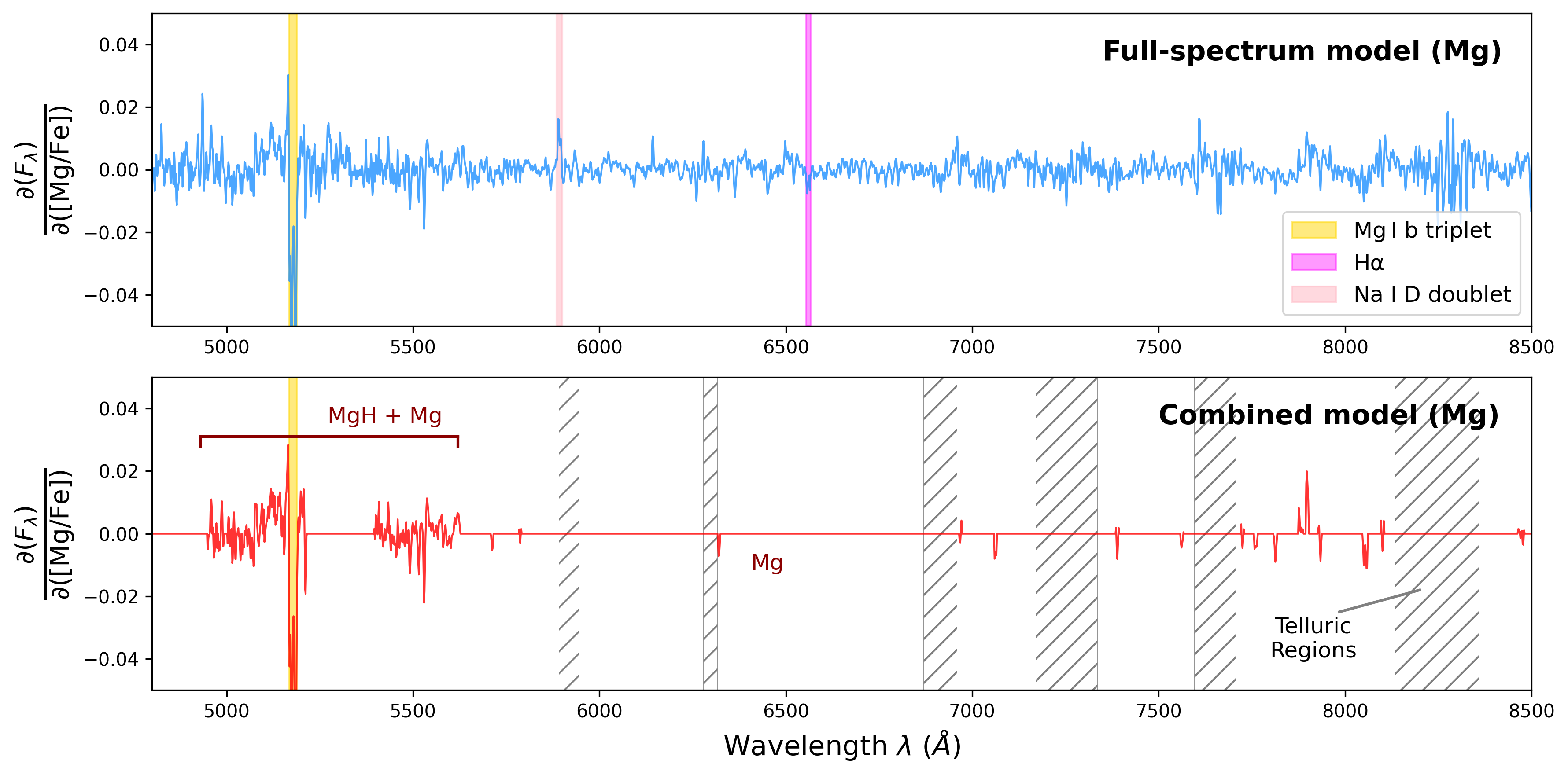}
    \caption{Coefficients ($\Theta_\lambda$) of the Full-spectrum model (upper panel) and Combined model (lower panel) that \textit{The Cannon} uses to determine the [Mg/Fe] from the spectra. The y-axis represent the variability of the flux as a function of Mg variability. The hatched regions are wavelength windows affected by telluric contamination. Additionally, Mg and MgH features were added.
    }
    \label{fig: theta_2models}
\end{figure*}

Figure~\ref{fig: theta_2models} shows the linear coefficients used to estimate Mg as a function of wavelength for the Full-spectrum model (upper panel) and the Combined model (lower panel). The regions shaded in yellow, pink, and magenta show the locations of the Mg triplet, Na~\textsc{i} doublet, and H$_\alpha$ absorption lines. The magnitude of these coefficients describes the model’s sensitivity to each wavelength pixel when inferring each label (magnesium abundance in this case). Thus, the peaks seen in this figure allow us to visualize which spectral regions are (as determined by \textit{The Cannon}) important for each label estimation. For instance, from the lower panel we notice that \textit{The Cannon}, when estimating Mg, shows high sensitivity to spectral features associated with Mg and MgH (as expected). The upward features around $7900$ Å in the lower panel of this figure are likely associated with residual telluric contamination, as these regions contain relatively weak telluric features that are not removed by our telluric masking. However, we find that these features have a negligible impact on the inferred abundances. Further details on the differences between the coefficients of the Full-spectrum and Combined models are provided in Appendix \ref{app: theta amplitude}.

\subsubsection{Performance evaluation: stellar label estimation from LAMOST spectra}

During the training phase, \textit{The Cannon} learns the coefficients $\Theta_\lambda$ based on the known APOGEE labels, $\ell_n^A$, and the corresponding LAMOST spectra for each reference star $n$. A straightforward way to validate our models is to use these learned coefficients $\Theta_\lambda$ to infer new labels, $\ell_n^L$, directly from the LAMOST spectra, $f_{n\lambda}^L$. This allows us to assess model accuracy by comparing the inferred labels $\ell_n^L$ with the true APOGEE values $\ell_n^A$.

Figure~\ref{fig: 1:1 APOGEEvsCANNON} shows the results of take-25\%-out cross-validation (see Section~\ref{subsec: cannon-censoring-validation}) for the Combined model, comparing the labels predicted by \textit{The Cannon} from LAMOST spectra against the corresponding APOGEE reference labels. Each panel reports the bias (median offset) and scatter (standard deviation) of the residuals (predicted $-$ reference). 
Table~\ref{tab:validation_models} summarizes the bias and scatter for all four models in predicting the abundances of [Mg/Fe], [Mn/Fe], [Al/Fe], [C/Fe], and [N/Fe]. The Mask column reports the fraction of spectral pixels used by the combined model for each element, i.e., the portion of the spectrum contributing to the estimation of each label ({21\% for Mg, 10\% for Mn, 2\% for Al, and 59\% for C and N).

\begin{figure*}
    \centering
    \includegraphics[width=1\linewidth]{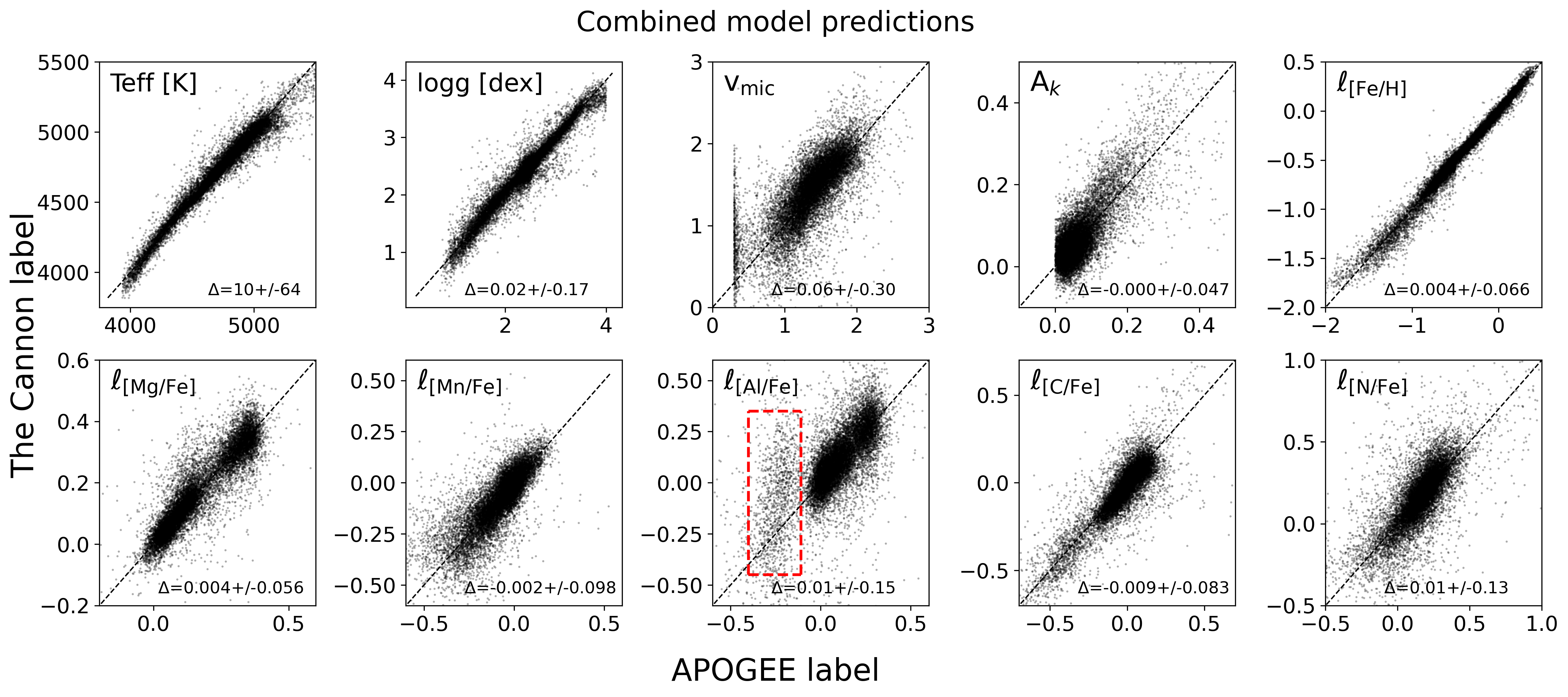}
    \caption{Cross-validation of the Combined Model build using 26,989 reference objects using LAMOST spectra and APOGEE labels. The y-axis shows the prediction from \textit{The Cannon} for each stellar label on the training set while the x-axis shows the (true) APOGEE values. The metrics reported are bias and RMS (root-mean-square) scatter around the 1:1 line (segmented line). Stars with poor fits ($\chi^2 > 15000$, comprising 6\% of the sample) were excluded from the analysis. The red box highlights stars with [Al/Fe] < 0.11, where the model shows degraded performance.} 
    \label{fig: 1:1 APOGEEvsCANNON}
\end{figure*}

\begin{table*}
\centering
\caption{Bias and RMS scatter around the 1:1 line for the predictions of each model. 
The last column (Mask) reports the percentage of spectral pixels included in the mask for each label in the combined model. The values in bold show the model with the best precision.}
\begin{tabular}{lcccccc}
\hline \hline
Label & Full-spectrum model & Absorption model & Telluric model & Combined model & Mask / \% \\ \hline 
Teff & $~~3 \pm 67$ & $9 \pm 72$ & \boldmath{$4 \pm 61$} & $10 \pm 64$ \\ 
logg & $~0.00 \pm 0.17$ & $0.01 \pm 0.19$ & \boldmath{$0.00 \pm 0.15$} & $0.02 \pm 0.17$ \\ 
Vmic & $~0.03 \pm 0.33$ & $0.07 \pm 0.32$ & $0.03 \pm 0.32$ & \boldmath{$0.06 \pm 0.30$} \\ 
Ak & $-0.001 \pm 0.046~$ & $0.000 \pm 0.047$ & \boldmath{$0.000 \pm 0.046$} & $0.000 \pm 0.047$ \\ 
Fe & $~0.001 \pm 0.071$ & $0.003 \pm 0.081$ & \boldmath{$0.002 \pm 0.062$} & $0.004 \pm 0.066$ \\ 
Mg & $-0.004 \pm 0.077~$ & $0.004 \pm 0.063$ & $-0.003 \pm 0.065$ & \boldmath{$0.004 \pm 0.056$} & 21 \\ 
Mn & $-0.00 \pm 0.12~$ & $0.00 \pm 0.099$ & $-0.00 \pm 0.11$ & \boldmath{$-0.002 \pm 0.098$} & 10 \\ 
Al & $~~0.02 \pm 0.13$ & $0.01 \pm 0.15$ & \boldmath{$0.02 \pm 0.12$} & $0.01 \pm 0.15$ & 2 \\ 
C & $-0.008 \pm 0.099$ & $-0.009 \pm 0.094~$ & $-0.008 \pm 0.085~$ & \boldmath{$-0.009 \pm 0.083~$} & 59 \\ 
N & $~~0.00 \pm 0.16$ & $0.01 \pm 0.15$ & $-0.00 \pm 0.14$ & \boldmath{$0.01 \pm 0.13$} & 59\\
\hline \hline
\end{tabular}
\label{tab:validation_models}
\end{table*}

\subsection{Analysis of Sculptor MUSE spectra}

\subsubsection{Continuum normalization and radial velocity correction}
\label{subsec:sculptor_preprocessing}

As with the training set, the raw MUSE spectra of the Sculptor RGB stars must be preprocessed before they can be analyzed with \textit{The Cannon}. We apply the same continuum normalization procedure described in Section~\ref{subsec: LAMOST}: dividing each spectrum by a Gaussian-smoothed version of itself (kernel width $\sigma = 50$ Å) to remove the underlying stellar continuum shape.

Unlike the LAMOST data, the MUSE spectra do not include pre-computed redshift or radial velocity information. We therefore measure radial velocities for the Sculptor stars through cross-correlation with the $0^{th}$ coefficient of our trained Full-spectrum \textit{The Cannon} model. This $0^{th}$ coefficient, $\theta_0(\lambda)$, represents the mean model spectrum for stars in the training set and serves as a template at rest frame. For the radial velocity measurements, we use the template restricted to the wavelength range covered by the MUSE observations. By measuring the wavelength shift between this template and each Sculptor spectrum, we derive the radial velocity of each star.

We compute radial velocities for the 42 selected Sculptor RGB stars (Section \ref{subsec: museHST_details}) using the \texttt{crosscorrRV} function from the \texttt{PyAstronomy} package in Doppler mode, exploring a velocity range of $-250$ to $+250\,\mathrm{km\,s^{-1}}$ with a sampling of $0.1\,\mathrm{km\,s^{-1}}$. This routine first interpolates the observed spectrum onto the wavelength grid of the template, then computes the cross-correlation function (CCF) across the specified velocity range. We fit a fifth-order polynomial to the resulting CCF, and adopt the peak of this fitted function as the stellar radial velocity. 

\begin{figure}
    \centering
    \includegraphics[width=0.8\linewidth]{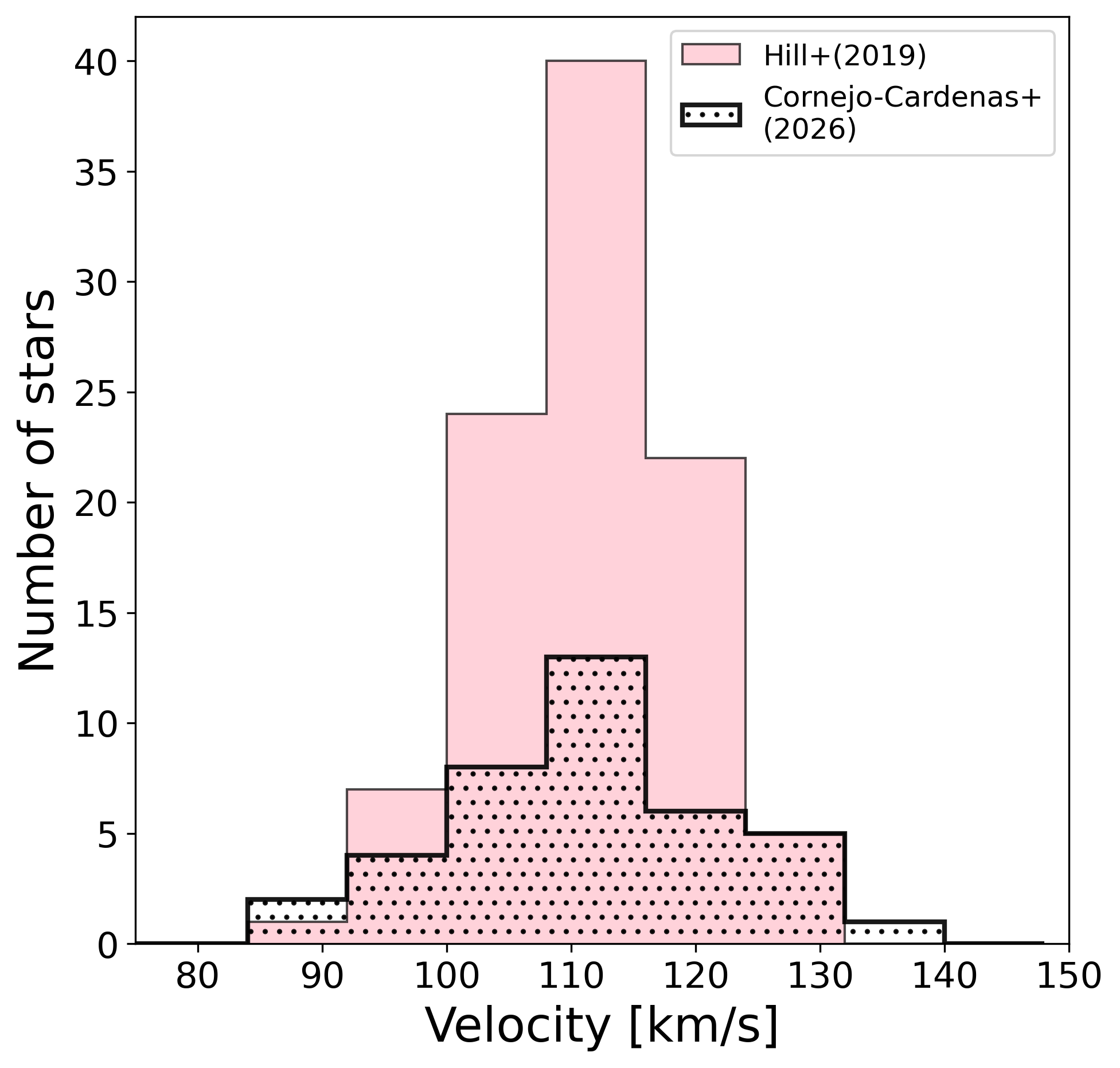}
    \caption{Radial velocity distribution of the 39 RGB stars in Sculptor derived from cross-correlation with the \textit{The Cannon} model template.}
    \label{fig: RV_scl}
\end{figure}

To estimate the uncertainties of the radial velocity measurements, we performed a Monte Carlo analysis by generating 10 noise realizations of each observed spectrum. For each realization, Gaussian noise was added to every pixel according to its flux uncertainty, and the radial velocity was re-measured using the same cross-correlation procedure described above. Pixels with artificially inflated uncertainties were excluded during this uncertainty estimation. This results in a median radial velocity uncertainty $\sigma_\text{measurement} = 4.8~\mathrm{km\,s^{-1}}$ per star for our RGB sample.

The resulting radial velocity distribution has a median value of $\langle v_r\rangle = 112.0~\mathrm{km\,s^{-1}}$. Three stars from the sample of 42 RGB stars are identified as lying beyond $2\sigma$ from the median radial velocity. To further assess their membership, we cross-matched our sample with Gaia DR3 and examined their proper motions. Gaia measurements are available for two of these three stars. Both stars have proper motions that lie beyond the $2\sigma$ range of the proper motion distribution measured for Gaia sources in the Sculptor field. Given that these stars show both radial velocities and proper motions that differ from the main distribution of our sample, we cannot confirm their membership in Sculptor and exclude them from the following analysis.

For the remaining 39 RGB stars, we obtain an observed radial velocity scatter of $\sigma_\text{spread}=11.0~\mathrm{km\,s^{-1}}$. Accounting for the median measurement uncertainty of $\sigma_{\rm measurement}=4.8~\mathrm{km\,s^{-1}}$, we estimate an intrinsic velocity dispersion of $\sigma_{\rm intrinsic}=\sqrt{\sigma_{\rm spread}^2-\sigma_{\rm measurement}^2}=9.9~\mathrm{km\,s^{-1}}$. This is similar with the intrinsic dispersion of $7.6~\mathrm{km\,s^{-1}}$ obtained from the measurements of \citet{Hill2019}, assuming a velocity scatter of $7.9~\mathrm{km\,s^{-1}}$ and a measurement uncertainty of $2.1~\mathrm{km\,s^{-1}}$ from \citet{Skuladottir2017}. The difference may reflect differences in the stellar samples and spatial regions of Sculptor probed by the different studies. Figure~\ref{fig: RV_scl} shows the resulting radial velocity distribution for our stars (dot-hatched histogram), compared with the measurements of \citet{Hill2019} (pink).

\subsubsection{Label inference for Sculptor stars} \label{subsec:sculptor_results}

Based on the cross-validation performance metrics presented in Table~\ref{tab:validation_models}, the Combined model achieves the lowest scatter (RMS) for most stellar labels compared to the other model variants (see Section~\ref{subsec:model_comparison} for more details). We therefore adopt the Combined model as our primary method for inferring stellar parameters and chemical abundances for the 39 RGB stars in Sculptor. Additionally, as a control, we apply the Full-spectrum model to the same Sculptor sample. 

The $\chi^2$ statistic quantifies the difference between the model spectrum and the observed spectrum, weighted by the measurement uncertainties at each pixel $\lambda$ as:

\begin{equation}\label{eq:chi2}
    \chi^2 = \sum_\lambda\frac{(f_{\rm model}(\lambda) - f_{\rm obs}(\lambda))^2}{\sigma^2(\lambda) + s(\lambda)^2}
\end{equation}

\noindent where $f_{\rm model}(\lambda)$ is the spectrum predicted by \textit{The Cannon} given the inferred stellar labels, $f_{\rm obs}(\lambda)$ is the observed MUSE spectrum, $\sigma(\lambda)$ is the (observed) flux uncertainty at each wavelength pixel, and $s(\lambda)$ is the intrinsic scatter of the model.

We report the reduced $\chi^2$, defined as $\chi^2_{\rm red} = \chi^2 / (N_{\rm pix} - L - 1)$, where $N_{\rm pix}$ is the number of valid spectral pixels and $L = 10$ is the number of inferred stellar labels ($T_{\rm eff}$, $\log g$, $v_{\rm mic}$, $A_K$, [Fe/H], [Mg/Fe], [Mn/Fe], [Al/Fe], [C/Fe], [N/Fe]). This quantity accounts for the number of fitted labels and provides a normalized measure of the model fit quality. For the calculation of $\chi^2_{\rm red}$, we exclude pixels with artificially inflated uncertainties, which correspond primarily to regions affected by telluric contamination. The Full-spectrum and Absorption models exclude approximately 130 pixels ($\sim5 \%$ of the spectrum), while the Telluric and Combined models exclude approximately 500 pixels ($\sim19\%$), including regions where uncertainties were increased to account for strong telluric features.

After this correction, all four models yield median reduced $\chi^2$ values close to unity, indicating that the inferred uncertainties are consistent with the observed residuals. Figures~\ref{fig: spectra_comparison_highsnr} and \ref{fig: spectra_comparison_lowsnr} show representative examples of the spectra generated by the Full-spectrum (blue) and Combined (red) models for two Sculptor stars with SNR = 37.76 (with $\chi^2_\text{red}(\rm full) = 0.97$ and $\chi^2_\text{red}(\rm combined) = 1.08$) and SNR = 11.69 (with $\chi^2_\text{red}(\rm full) = 0.81$ and $\chi^2_\text{red}(\rm combined) = 0.87$) respectively.

Figure~\ref{fig: uncertainties_vs_snr} shows the label uncertainties ($T_{\rm eff}$, $\log g$, [Fe/H], [Mg/Fe], [Mn/Fe], [C/Fe], [N/Fe]) as a function of SNR for the 39 resolved Sculptor stars, derived using the Combined model. Uncertainties were estimated as standard deviations of fit results from the 10 Monte Carlo spectrum realizations described in Section~\ref{subsec: cannon-censoring-validation}. These uncertainties represent the precision of the inferred labels due to random spectral noise and do not include possible systematic uncertainties related to the model or training set. Each track in this figure represents the running median of individual stellar uncertainties, with shaded regions indicating the standard deviation around this median. The vertical dashed gray line marks SNR = 25, dividing the sample into low-SNR (17 stars, 40\% of the sample) and high-SNR (22 stars, 60\%)} subsets. 

We exclude [Al/Fe] from the figure because the Combined model has known limitations in estimating this abundance, resulting in higher uncertainties than for the other elements. Its typical uncertainty decreases from $\sigma([\rm Al/Fe]) \sim 0.82$ dex in the low-SNR subset to $\sigma([\rm Al/Fe]) \sim 0.58$ dex in the high-SNR subset, remaining outside the y-axis range shown in the figure.

\begin{figure}
    \centering
    \includegraphics[width=1\linewidth]{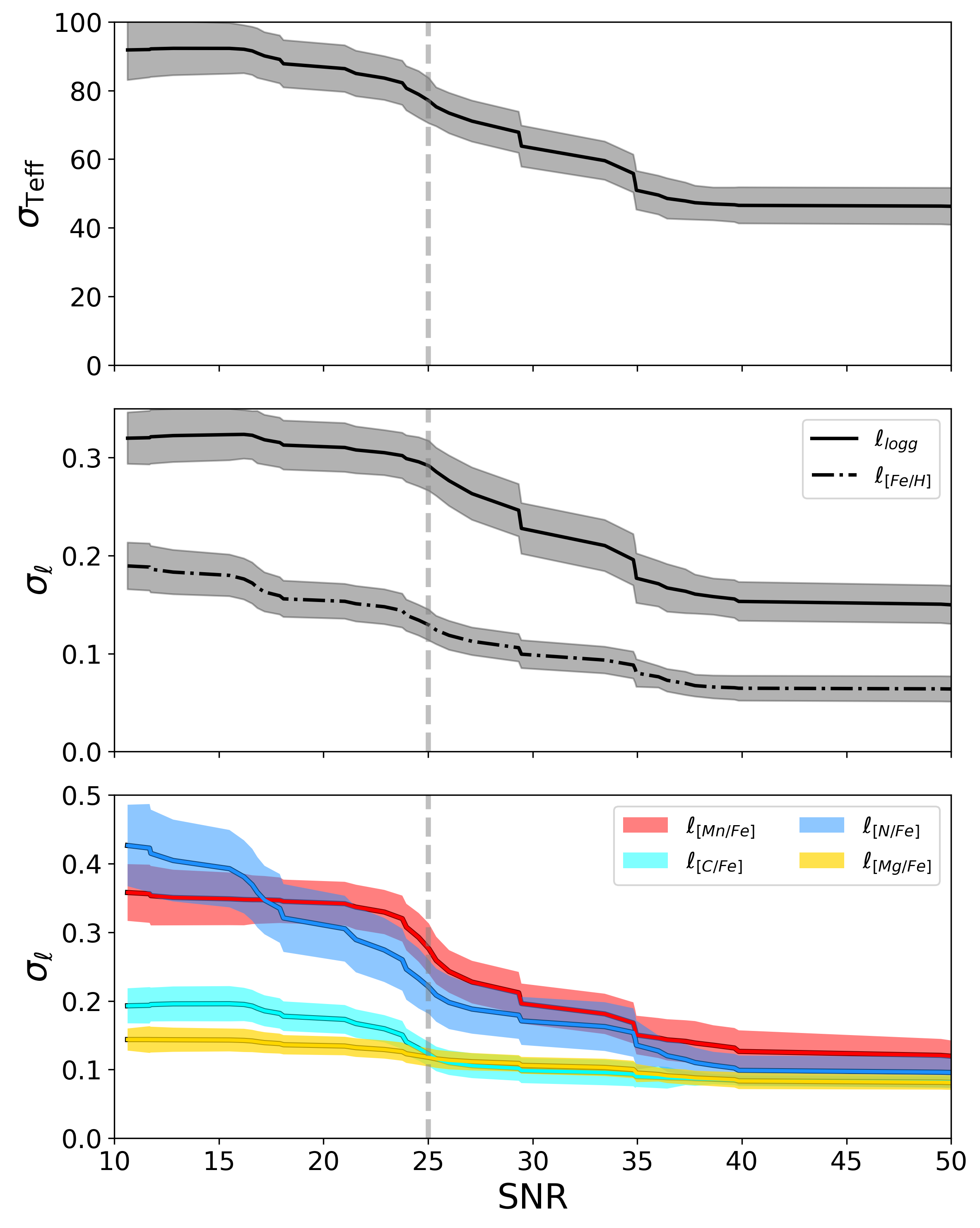}
    \caption{Label uncertainties as a function of SNR for the 39 Sculptor stars, estimated via Monte Carlo sampling (10 realizations per star). Tracks show running medians with shaded standard deviations. The vertical dashed line marks SNR = 25, separating low-SNR (40\%) and high-SNR (60\%) subsamples.}
    \label{fig: uncertainties_vs_snr}
\end{figure}

As expected, uncertainties decrease with increasing SNR for all labels. For stellar parameters, typical uncertainties are $\sigma(T_{\rm eff}) \sim 94$ K and $\sigma(\log g) \sim 0.33$ dex at low SNR, improving to $\sim 47$ K and $\sim 0.16$ dex at high SNR. Chemical abundance uncertainties range from $\sim 0.14$--$0.39$ dex (low SNR) to $\sim 0.065$--$0.14$ dex (high SNR). Iron shows the most significant variation, with $\sigma([\rm Fe/H])$ decreasing from $\sim 0.18$ dex to $\sim 0.065$ dex between the low- and high-SNR regimes, corresponding to a reduction by a factor of $\sim 2.8$.

These trends are generally consistent with those reported in \citetalias{Ness2025}. However, our Combined model achieves lower $T_{\rm eff}$ uncertainties: at SNR $\sim 10$, \citetalias{Ness2025} (using an equivalent to the Full-spectrum model of this work) shows a $\sigma(T_{\rm eff}) \sim 170$ K, compared to $\sim 92$ K in this work.

\begin{figure}
    \centering
    \includegraphics[width=1\linewidth]{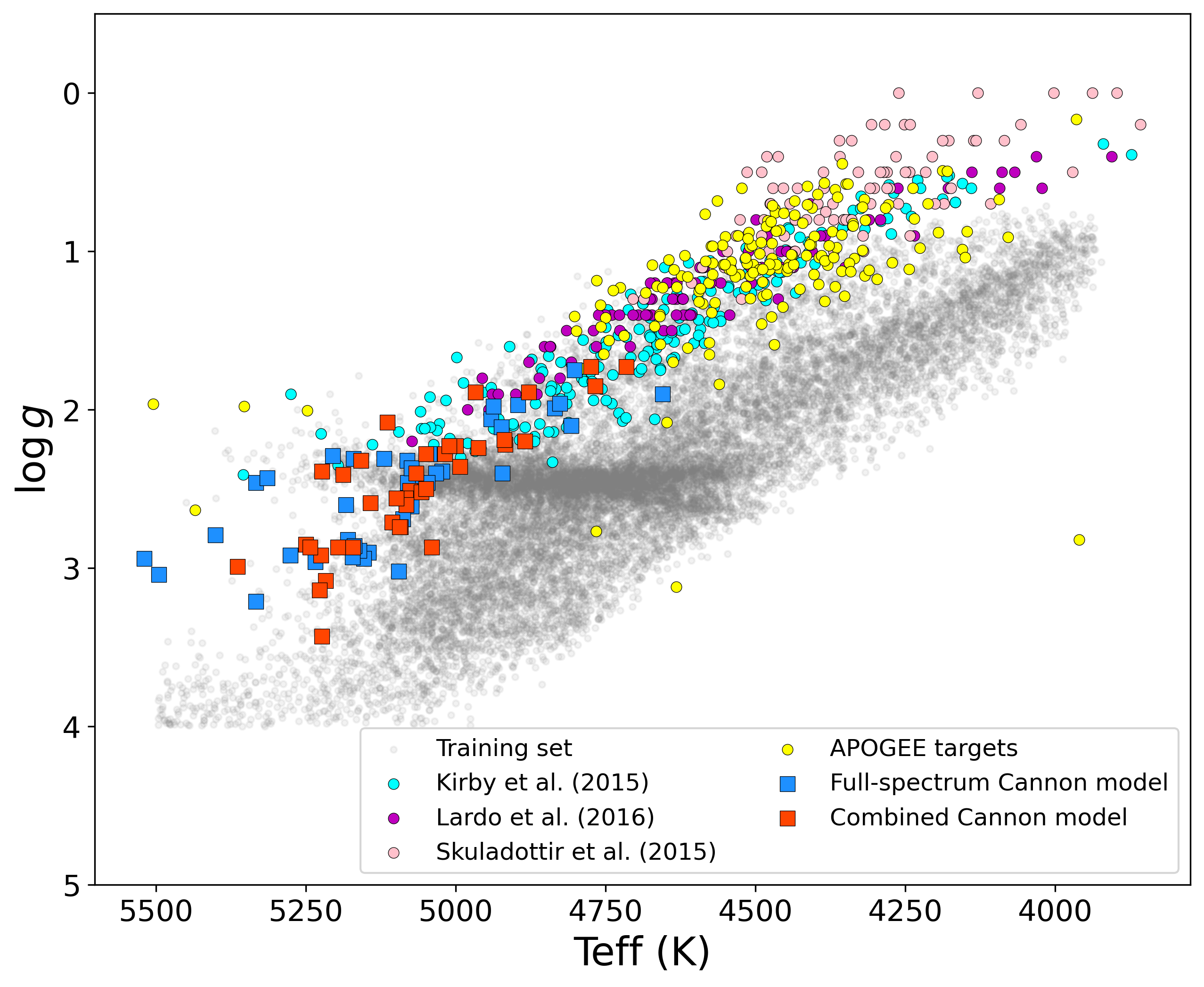}
    \caption{$T_{\rm eff}$--$\log g$ diagram for Sculptor RGB stars. Our predictions from full-spectrum (blue squares) and Combined (red squares) models compared to literature samples: \citet{Kirby2015} (cyan), \citet{Lardo2016} (magenta), \citet{Skuladottir2015} (pink). In gray is the training set (MW stars) and in yellow are the APOGEE Sculptor field observations.}
    \label{fig: logg_teff}
\end{figure}

\begin{figure}
    \centering
    \includegraphics[width=1\linewidth]{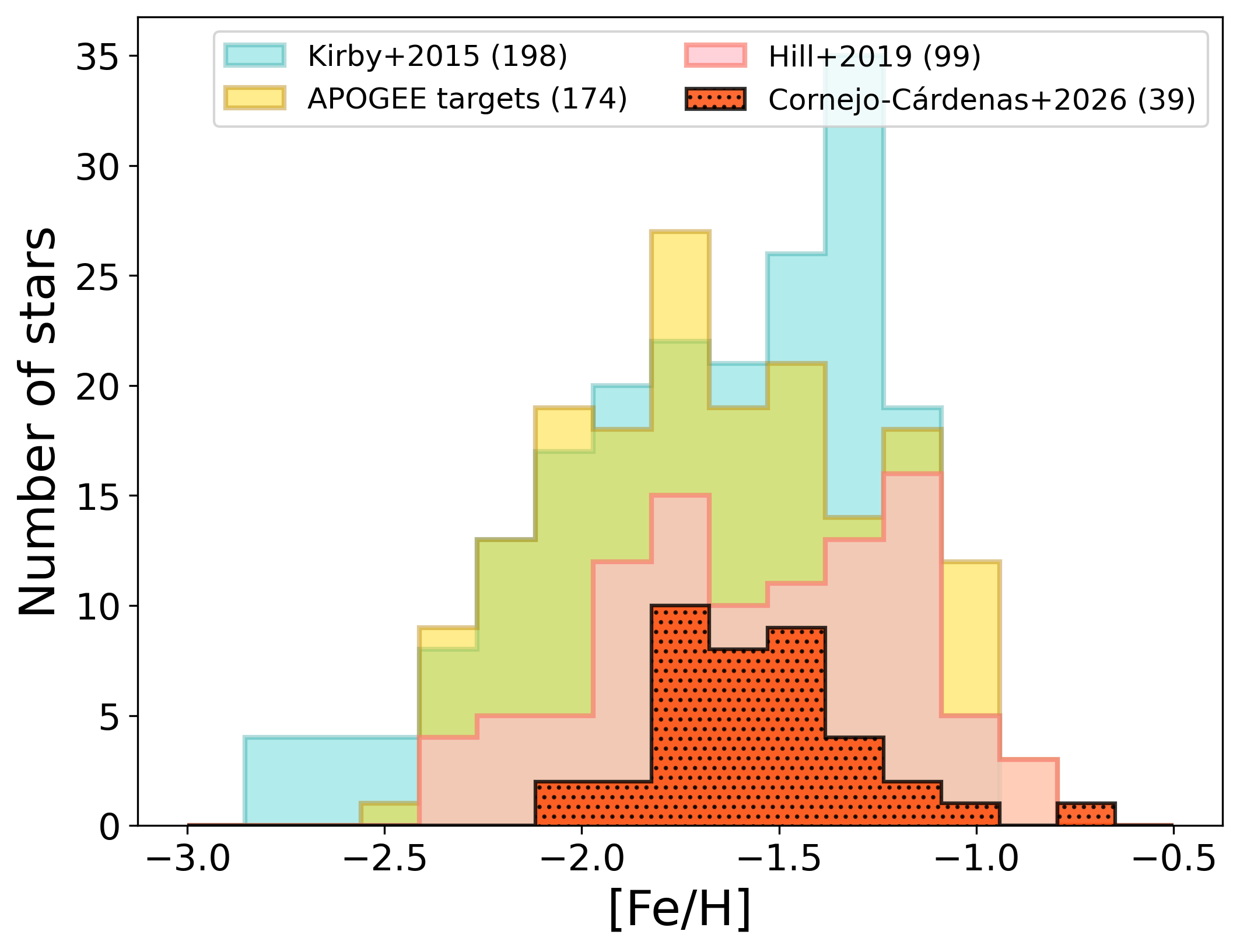}
    \caption{A comparison of the inferred [Fe/H] label for our Combined model (orange dot-shaded distribution) with that of medium resolution analyses of \protect\cite{Kirby2015} and high resolution spectroscopy from \protect\cite{Hill2019}. The number of Sculptor stars used in each study is indicated in parenthesis.}
    \label{fig: histogram_feh}
\end{figure}

\begin{figure*}
    \centering
    \includegraphics[width=1\linewidth]{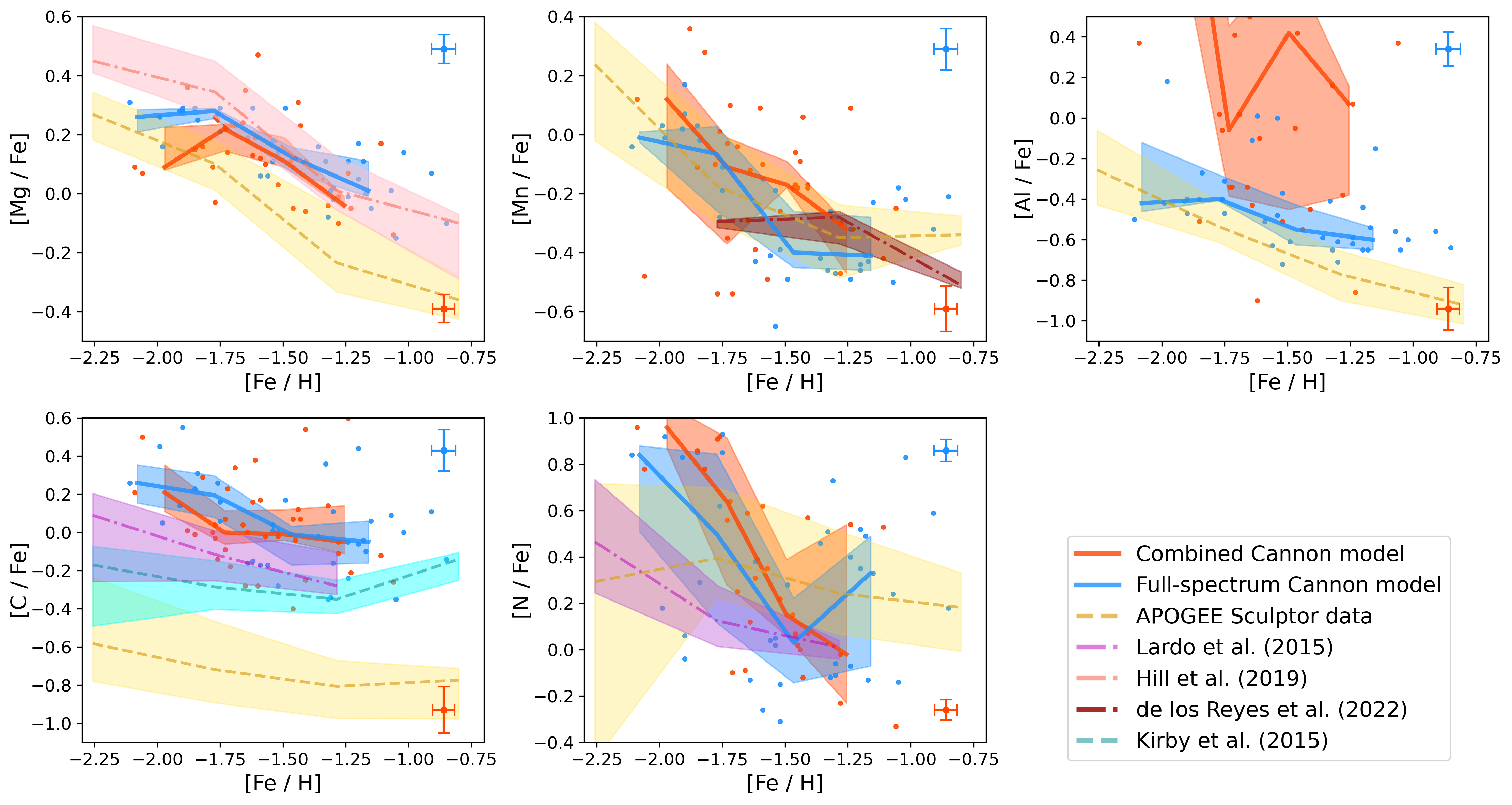}
    \caption{[X/Fe] as a function of [Fe/H] for 39 stars in the Sculptor dwarf galaxy, derived using the Full-spectrum (blue solid line) and Combined (red solid line) models (considering X=Mg, Mn, Al, C and N). The tracks show median values in [Fe/H] bins, with shaded areas marking the 25th–75th percentiles. Literature data are shown for comparison: \citet[][magenta dash–dotted line]{Lardo2016}, \citet[][pink dash–dotted line]{Hill2019}, \citet[][cyan dash line]{Kirby2015}, \citet[][brown dash–dotted line]{delosReyes2022}, and APOGEE observations in the Sculptor field (yellow dash line). Crosses indicate the median abundance uncertainties from \textit{The Cannon} for each model.}
    \label{fig: Result: x/fe vs fe/h}
\end{figure*}

\section{Discussion } \label{sec:discussion}

\subsection{Training set selection and coverage} \label{subsec:training_set_coverage}

One key difference between our approach and ACACIAS \citetalias{Ness2025} is the evolutionary state coverage of the training set. In \citetalias{Ness2025}, RGB stars were selected using a surface gravity cut of $\log g < 2.5$, focusing on the most evolved giants. In contrast, we adopt a less restrictive criterion of $\log g < 4.0$ (Section~\ref{subsec: APOGEE}). This decision was motivated by initial tests showing that a $\log g < 2.5$ cut led to under-representation of certain parameter combinations in the training set, forcing \textit{The Cannon} to extrapolate rather than interpolate when predicting labels for some less luminous Sculptor stars.

Figure~\ref{fig: logg_teff} illustrates this issue in the $\log g$--$T_{\rm eff}$ plane. Gray points show the training set distribution (Milky Way stars), with a prominent overdensity at $\log g \sim 2.5$ corresponding to red clump stars. Blue and red squares represent the \textit{The Cannon} predictions for Sculptor stars using the Full-spectrum and Combined models, respectively. Approximately 50\% of the 39 Sculptor RGB stars have predicted $\log g < 2.5$, placing them outside the parameter space densely sampled by a training set restricted to $\log g < 2.5$. Extending the training set to $\log g < 4.0$ ensures that the Sculptor stars remain within the $\log g$--$T_{\rm eff}$ parameter space covered by the training set, allowing \textit{The Cannon} to interpolate rather than extrapolate.

Compared to previous spectroscopic studies of Sculptor \citep{Kirby2015,Lardo2016,Skuladottir2015}, our sample occupies a parameter space with higher $\log g$ and $T_{\rm eff}$. This suggests we are observing stars at earlier evolutionary stages along the RGB, whereas previous studies preferentially targeted more evolved, brighter giants. This difference in evolutionary phase may introduce systematic effects when comparing abundances between studies, particularly for elements sensitive to stellar evolution (e.g., C, N; see Sec.~\ref{subsubsec: lit_comparison} for more details).

The expanded $\log g$ range also increases the number of metal-poor stars in the training set. As shown by the metallicity distribution of the training set in Figure~\ref{fig: feh_comparison} (orange histogram), the sample is sparsely populated below [Fe/H]~$=-1$, with a sharp decline in density at [Fe/H]~$< -1.5$. The mean metallicity of the Sculptor sample ($\langle$[Fe/H]$\rangle \sim -1.6$; see Figure~\ref{fig: histogram_feh}) therefore lies in a regime where the training-set of \citetalias{Ness2025} coverage is intrinsically low.

However, relaxing the $\log g$ cut from 2.5 to 4.0, combined with the metallicity-dependent SNR selection described in Section~4.1.1, increased the number of metal-poor giants included in the training set. In \citetalias{Ness2025}, only $\sim$5\% of the training stars had [Fe/H]~$< -1$, whereas in the expanded training set this fraction increases to $\sim$11\%, improving the overlap between the Milky Way–based training distribution and the metallicities of our Sculptor targets. Nonetheless, future extensions of the ACACIAS program to more metal-poor galaxies will require supplementing the training set with additional extremely metal-poor giants.

\subsection{Model performance comparison: which model works best and why?}
\label{subsec:model_comparison}

Table~\ref{tab:validation_models} summarizes the performance of each model in predicting chemical abundances compared to the APOGEE reference labels, as measured through cross-validation on the training set. The table reports two metrics: bias (median offset between predicted and reference values from APOGEE) and RMS scatter (root-mean-square of residuals). All four models (Full-spectrum, Absorption, Telluric, and Combined) show biases close to zero for all labels, indicating no significant systematic offsets with respect to APOGEE. Additionally, the distribution of individual predictions compared with their expected value from APOGEE is shown in Figure \ref{fig: 1:1 APOGEEvsCANNON} for the Combined model. 

From Table~\ref{tab:validation_models}, we observe a progressive decrease in RMS for Mg, Mn, C, and N when moving from the Full-spectrum model through the Absorption and Telluric  models to the Combined model. This pattern indicates that both physically motivated wavelength selection and telluric masking contribute to reduced scatter in abundance predictions. Therefore, the Combined model, which incorporates both strategies, achieves the lowest RMS for these elements.

Quantitatively, Mg shows the most significant reduction among the other elements, decreasing its RMS by 27\% when moving from the Full-spectrum model to the Combined model. While, the elements Mn, C and N show an RMS reduction of 18\%, 16\% and 19\%, respectively, when using the Combined model. Aluminium shows a different pattern, with RMS increasing from 0.13 dex (Full-spectrum) to 0.15 dex (Combined). We interpret this as evidence that \textit{The Cannon} relies on spectral information beyond the known Al absorption lines when predicting [Al/Fe] in the Full-spectrum model. Specifically, the Na~\textsc{i} doublet at 5890/5896 Å, while not an aluminium feature, is correlated with aluminium abundances in the training set, due to their similar nucleosynthetic origins \citep{Kobayashi2020}. By restricting the Combined model to only the designated Al features, we exclude this informative spectral region, resulting in higher RMS. 

Furthermore, beyond the correlation with the Na~\textsc{i} doublet, a more fundamental consideration is whether the signal in the designated Al pixels exceeds the noise level. The Combined model restricts \textit{The Cannon} to only $\sim 2\%$ of the spectral pixels (see the Mask column in Table~\ref{tab:validation_models}), and the utility of this restricted set depends critically on whether the flux in those pixels carries sufficient signal above the noise. This condition is sensitive to both the metallicity of the star and the SNR of the spectra, as we discuss further in Section~\ref{subsubsec: lit_comparison}.

Comparing the Combined and Telluric models, we find that their performance metrics are generally similar across most labels, with differences not exceeding 14\%. The Mg, Mn, C, and N abundances show variations of 14\%,11\%, 2\%, and 7\% respectively when transitioning from the Telluric to the Combined model. Similarly, the fundamental stellar parameters ($T_{\rm eff}$, $\log g$, and [Fe/H]) show RMS differences of 5\%, 12\%, and 6\% respectively between the two models. These variations indicate that there is no significant difference in the performance of both models for most labels. However, aluminium shows a more significant difference: the Telluric model achieves 20\% lower RMS relative to the Combined model (and 8\% lower RMS relative to the Full-spectrum model). This is consistent with \textit{The Cannon} extracting information from other spectral features to estimate [Al/Fe], and the telluric masking strategy effectively isolate these diagnostic features from telluric contamination.

For the stellar parameters ($T_{\rm eff}$, $\log g$, and [Fe/H]), the Telluric model shows the lowest RMS values, despite the overall similarity with the Combined model performance. This difference is expected, as temperature, surface gravity, and overall metallicity affect the entire spectrum through continuum shape, pressure-broadening effects, and numerous Fe absorption lines distributed across the wavelength range. The Telluric model benefits from masking poorly corrected telluric features, which reduces systematic biases in the full-spectrum fitting and allows these global stellar properties to be recovered more precisely from the uncontaminated spectral regions.

It is worth noting that these metrics were estimated using stars from the training set, whose LAMOST spectra have a median SNR of $\sim 200$. While this provides a reliable reference for model performance in the high-SNR regime, it may not fully reflect the performance expected for the MUSE spectra of Sculptor, which have a median SNR $\sim 25$. To assess whether the relative performance of the models remains consistent at lower SNR, in Appendix~\ref{app: degraded_spectra} we repeat the cross-validation on validation spectra artificially degraded to the typical SNR of our MUSE sample. The results, summarised in Table~\ref{tab:degraded_validation_models}, show that the different models deliver comparable recovery across the different labels.

Additionally, Figure \ref{fig: 1:1 APOGEEvsCANNON} shows that predictions of [Al/Fe] < -0.11 for the Combined model, highlighted by the red box, exhibit higher dispersion (0.29 dex) compared to those with [Al/Fe] > -0.11 (0.15 dex). This increased dispersion at lower aluminium abundances is attributed to the lower SNR and line depths in these spectra, which degrades prediction precision. This trend is observed for all models: for [Al/Fe] < -0.11, the Combined and Absorption models exhibit the highest scatter (0.29 dex for both of them), while the Full-spectrum and Telluric models show achieve better precision (0.22 and 0.18 dex respectively). These values are consistent with the RMS values presented in Table \ref{tab:validation_models}, further supporting the conclusion that the Telluric model provides the most precise [Al/Fe] estimates.

For completeness, Figure \ref{fig: 1:1 APOGEEvsCANNON} also shows the predictions of V$\rm_{mic}$ and Ak, which were included in the training set to improve the flexibility of the model for line shapes (V$\rm_{mic}$) and external influences such as interstellar absorption ($A_K$) as done in previous analyses \citep[e.g.][]{Buder2018}. 

\subsection{Which model should you choose? Correlation vs causality}\label{subsec: which model choose?}

The choice of model depends on the scientific question being addressed and whether one prioritizes correlation-based predictions or physically interpretable (causal) relationships.

The Full-spectrum and Telluric models use information from the entire spectrum to predict stellar labels, learning correlations between spectral flux and labels without imposing physical constraints on which wavelengths should matter for which labels. This approach maximizes the information available to the model and can achieve low RMS by exploiting all correlations in the data, including indirect ones. However, it provides limited insight into which spectral features drive each prediction. The Combined model, in contrast, is physically motivated, as it restricts \textit{The Cannon} to spectral regions where we expect each element to have strong absorption features based on atomic physics. This tests whether the model can learn the correct abundance patterns from the physically relevant spectral information alone (i.e., causality rather than just correlation). When this approach achieves comparable or lower RMS than the Full-spectrum model (as it does for Mg, Mn, C, and N) it provides evidence that the abundance predictions are driven by the expected physical mechanisms rather than internal correlations.

Based on the RMS performance metrics, we recommend the following model choices:
\begin{itemize}[leftmargin=2em,labelwidth=2em]
    \item Based on the overall performance metrics presented in Table 1, we recommend the Combined model for determining most labels, as it provides the highest precision across Mg, Mn, C, and N abundances, with differences relative to the Telluric model remaining within 14\%. However, aluminum should be excluded from this recommendation.

    \item For precise [Al/Fe] abundance determinations, we recommend the Telluric model, which achieves 25\% and 8\% better precision than the Combined model and Full-spectrum model for this element. This is because these models can also use flux information from the Na~\textsc{i} doublet. However, because it is not based exclusively on Al absorption features, this prediction should be interpreted as an \textit{Al-like} abundance (i.e., $l_{\mathrm{[Al/Fe]}}$), particularly for stellar populations that may not follow the same abundance correlations as the training set.
    
\end{itemize}

It is worth noting that the SNR of the test spectra may also influence the choice of model. As shown in Appendix~\ref{app: degraded_spectra}, our recommendations remain broadly consistent when evaluated at a median SNR $\sim 25$, reflecting the typical quality of our Sculptor MUSE spectra. At lower SNR regimes, however, the relative performance of the models may change, particularly for models that rely on a restricted set of spectral pixels, and we encourage caution when extrapolating these recommendations beyond this range.

For our analysis of Sculptor RGB stars, we adopted the Combined model, as it showed the best RMS performance for the abundances of [Mg/Fe], [Mn/Fe], [C/Fe], and [N/Fe]. In addition, we will also use the Full-spectrum model (equivalent to the one used in \citetalias{Ness2025}) as a control, which will allow us to study the correlation behind the inferences.

\subsection{Sculptor abundances}\label{subsec: sculptor_abundances}
\subsubsection{Comparison with literature} \label{subsubsec: lit_comparison}

Figure~\ref{fig: Result: x/fe vs fe/h} compares our abundance measurements for Sculptor ([X/Fe] vs. [Fe/H]) from the Full-spectrum (blue) and Combined (red) models with values from the literature. Each track shows the median of [X/Fe] in bins of [Fe/H], with shaded regions representing the 25th and 75th percentiles within each bin. For our models, each bin is required to contain at least four stars. Individual points show the measurements for each star, while error bars indicate the median internal uncertainty from \textit{The Cannon} for each model. These uncertainties are computed as the square root of the diagonal elements of the covariance matrix returned by \textit{The Cannon} during the $\chi^2$ minimization, and represent the uncertainties associated with the label-fitting procedure. It is important to note that these internal uncertainties reflect the fitting uncertainties from \textit{The Cannon}'s $\chi^2$ minimization and are distinct from the uncertainties estimated via Monte Carlo sampling (Figure~\ref{fig: uncertainties_vs_snr}). The internal uncertainties are typically smaller and reflect how well the model reproduces the observed spectrum.

From this figure, we observe generally good agreement between the literature trends and our model predictions for most elements. For example, we find decreasing trends in [Mg/Fe], [Mn/Fe], and [C/Fe] with increasing [Fe/H], consistent with the distributions of APOGEE targets (see also  \citeauthor{Shetrone2026} \citeyear{Shetrone2026}) and measurements from \citet{Hill2019}, \citet{delosReyes2022}, \citet{Lardo2016}, and \citet{Kirby2015}. However, we note some tension between our tracks and those from the literature, which may arise from two factors. 

First, the tracks in this figure show only the median values in [Fe/H] bins to provide a clearer visualization of overall trends. When examining individual measurements (Figure~\ref{fig: sculptor_individual_scatter}), our predictions generally fall within the range of literature values. Second, our sample is concentrated toward the center of Sculptor, whereas the literature samples, including those from \citet{Hill2019}, \citet{Kirby2015}, \citet{Lardo2016}, and APOGEE, are more spatially distributed across the galaxy. If radial abundance gradients exist in Sculptor \citep{Tolstoy2004}, this spatial difference could introduce systematic offsets between our measurements and those from the literature, particularly for elements sensitive to star formation history.

The [Mg/Fe] trend in Figure ~\ref{fig: Result: x/fe vs fe/h} reflects Sculptor's chemical evolution: at early times (low [Fe/H]), $\alpha$-elements like Mg are preferentially produced by Type II supernovae from massive, short-lived stars, resulting in enhanced [$\alpha$/Fe] ratios. As Sculptor evolved and [Fe/H] increased, Type Ia supernovae, which produce more Fe relative to $\alpha$ elements and have longer delay times, began to contribute significantly to the chemical enrichment, causing [$\alpha$/Fe] to decline. 

Our [Mn/Fe] predictions follow a consistent pattern with previous observations from APOGEE and \citet{delosReyes2022}. Both the Full-spectrum and Combined models produce similar [Mn/Fe] tracks, indicating that \textit{The Cannon} is primarily extracting Mn abundance information from the designated Mn absorption features.

\begin{figure*}
    \centering
    \includegraphics[width=0.8\linewidth]{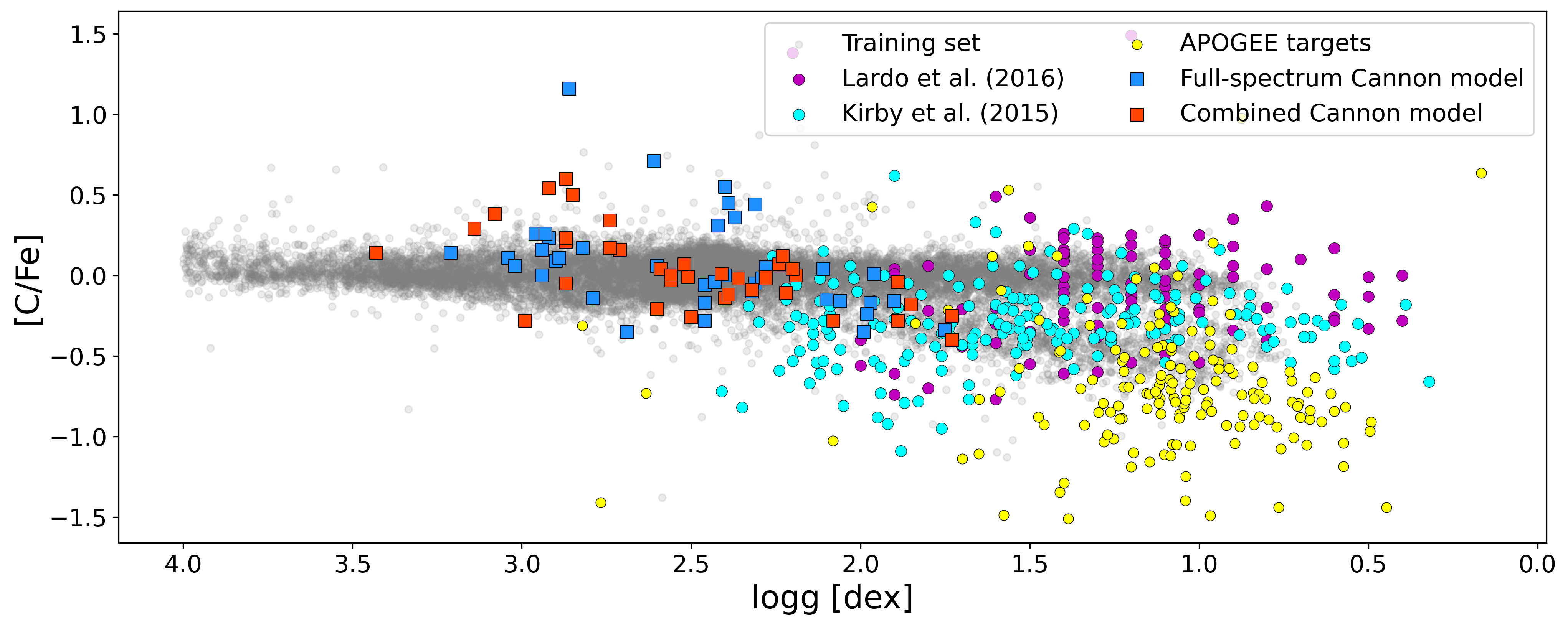}
    \caption{[C/Fe] abundance as a function of log$g$ for Sculptor RGB stars. Our predictions are shown as blue (Full-spectrum model) and red (Combined model) squares, while literature values are represented by cyan, magenta, and yellow circles (\citealt{Kirby2015}; \citealt{Lardo2016}; and APOGEE targets). For comparison, the gray dots correspond to the training set, composed of Milky Way RGB stars.}
    \label{fig: C_logg}
\end{figure*}

For carbon, we find that our models predict [C/Fe] abundances higher than values previously reported in the literature. This discrepancy likely arises because our Sculptor giant star sample occupies a different region of evolutionary track region compared to previous studies. Carbon is particularly sensitive to stellar evolution, as its surface abundance changes during dredge-up episodes along the RGB \citep{Placco2014}. Consequently, evolutionary corrections must be applied based on metallicity and surface gravity to recover the initial stellar carbon abundance.

Figure~\ref{fig: C_logg} shows the distribution of [C/Fe] as a function of $\log g$ for the training set (gray points, Milky Way RGB stars), literature measurements in Sculptor (cyan: \citealt{Kirby2015}; magenta: \citealt{Lardo2016}; yellow: APOGEE targets), and our predictions for the 39 Sculptor stars (blue squares: Full-spectrum model; red squares: Combined model). Our Sculptor sample occupies higher $\log g$ values than most literature samples, particularly those from APOGEE and \citet{Lardo2016}. This indicates that we are observing stars at earlier evolutionary stages, where carbon has undergone less depletion through mixing processes.

The Combined model predictions of [C/Fe] show better agreement with the overall trend established by literature values than the Full-spectrum model. This suggests that the carbon absorption features selected in our Combined model are good diagnostics of carbon abundance, and masking telluric-contaminated regions reduces systematic uncertainties associated with atmospheric absorption correction.

Figure~\ref{fig: C_logg} also shows that our Sculptor stars fall well within the $\log g$ range covered by our training set (gray points), confirming that \textit{The Cannon} is interpolating rather than extrapolating. This interpolation regime was the primary motivation for extending our training set to higher $\log g$ values compared to \citetalias{Ness2025}, ensuring robust abundance predictions for stars across a broader range of evolutionary states.

For nitrogen, we observe large scatter in [N/Fe] for both our model predictions and literature measurements (see Figures~\ref{fig: Result: x/fe vs fe/h} and \ref{fig: sculptor_individual_scatter}). Despite this scatter, our [N/Fe] estimates fall within the range established by previous studies, indicating that our models capture the overall nitrogen abundance distribution in Sculptor. 

Aluminium presents a more complex case. The Full-spectrum model recovers the APOGEE trend reasonably well (with a median of $\sim -0.50$ dex), whereas the Combined model systematically over-predicts aluminium abundances (with a median value of $\sim 0.07$ dex) compared to APOGEE measurements. As discussed in Section~\ref{subsec:model_comparison}, the increase in RMS from 0.13 to 0.15 dex when restricting the model to designated Al features indicates that these lines carry insufficient information at the SNR of the MUSE spectra. In the Full-spectrum model, \textit{The Cannon} can instead exploit correlated spectral features elsewhere in the spectrum, such as the Na~\textsc{i} doublet at 5890/5896 Å, which traces similar nucleosynthetic enrichment channels. When this additional information is excluded, the model becomes dominated by noise in the weak Al features, leading to unreliable [Al/Fe] predictions. As discussed above, the Full-spectrum model prediction should be interpreted as an $\ell_\text{[Al/Fe]}$ label rather than a direct measurement of [Al/Fe], as it incorporates correlated chemical information from the training set. 

\begin{figure*}
    \centering
    \includegraphics[width=1\linewidth]{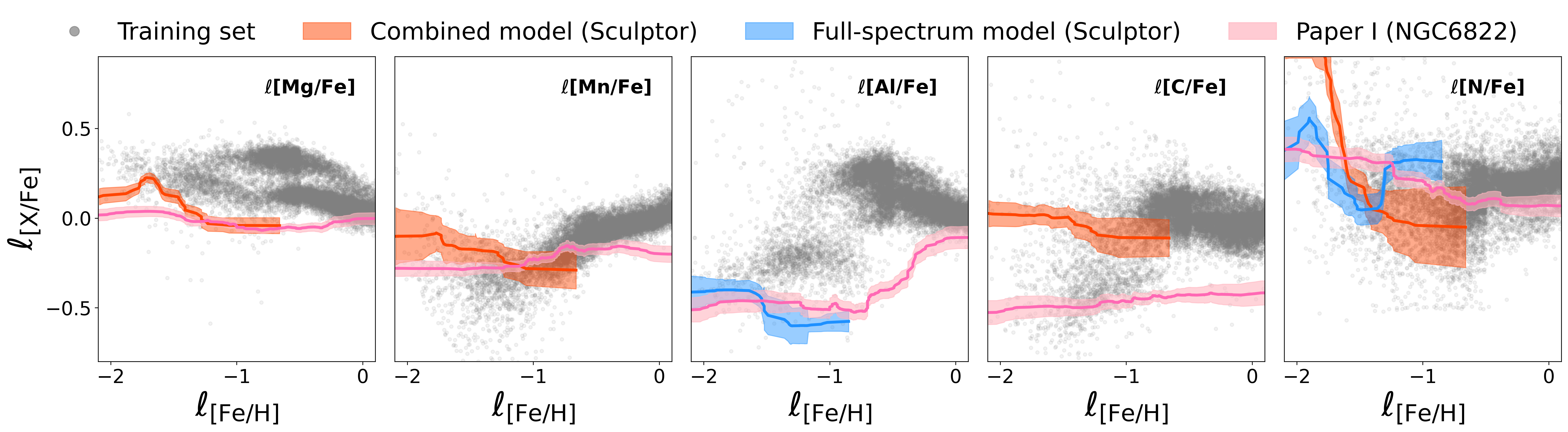}
    \caption{Element abundance distributions for 39 stars in the Sculptor dwarf galaxy derived using the Combined (red) and Full-spectrum (blue) models. For comparison, the training set labels are shown in grayscale and and the trend obtained in \citetalias{Ness2025} for the galaxy NGC 6822 is shown in pink.}
    \label{fig: comparison_ngc6822}
\end{figure*}

From Figure~\ref{fig: Result: x/fe vs fe/h}, we also observe that, excluding aluminium, our model predictions generally show larger discrepancies at low metallicities ([Fe/H] $\lesssim -1.6$). This behaviour likely reflects two factors. First, metal-poor stars remain under-represented in our training set despite our optimization efforts: only 11\% of training stars have [Fe/H] $< -1.0$ dex, and the fraction decreases further at [Fe/H] $< -1.6$ dex. Second, the metal-poor Sculptor stars tend to have lower spectral quality, with a median SNR of 24.0 compared to SNR $\sim 34.1$ for stars with [Fe/H] $\gtrsim -1.6$ dex. The combination of sparse training coverage and lower SNR poses particular challenges for the Combined model, which relies exclusively on designated absorption features for each element. 

To further investigate the origin of these discrepancies at low metallicities, we examined the signal-to-noise characteristics of the spectral features used by the Combined model. We divided the Sculptor stellar sample into metal-rich ($[\rm Fe/H] > -1.6$ dex, 48.7\% of the sample) and metal-poor ($[\rm Fe/H] < -1.6$ dex, 51.3\% of the sample) groups, and measured the fraction of pixels within the designated absorption windows for each element that satisfy the signal criterion $(1-\mathrm{flux})/\sigma > 1$. Metal-poor stars consistently exhibit fewer pixels meeting this threshold across all elements, indicating a systematic loss of usable spectral information at lower metallicities.

The magnitude of this pixel loss is comparable for Mg, Al, C, and N, with reductions of 12.4\%, 11.4\%, 13.2\%, and 13.2\%, respectively, when moving from the metal-rich to the metal-poor regime. Manganese shows a smaller reduction of 6.1\%. Although the relative pixel loss is comparable for most elements, aluminium relies on the smallest spectral window among the elements studied, using only 2\% of the total spectrum in the Combined model (see Mask column in Table~\ref{tab:validation_models}). Consequently, the smaller number of pixels available to constrain aluminium makes its abundance estimates particularly sensitive to the loss of spectral information at low metallicities, contributing to the poorer performance of the Combined model for aluminium compared with the other elements examined in this work.

The Combined model remains a viable option for abundance determinations across a range of metallicities. However, we recommend exercising caution when interpreting $[\rm Mg/Fe],~[\rm Al/Fe], ~[\rm C/Fe]$, and $[\rm N/Fe]$ estimates for metal-poor stars ($[\rm Fe/H] < -1.6$ dex) with low spectral quality (SNR $\lesssim 25$). In such cases, the Full-spectrum model may provide more robust estimates by leveraging information from the entire spectrum rather than relying exclusively on designated absorption features. Additionally, to validate abundance predictions for these stars, independent measurements from high-resolution spectroscopy obtained with next-generation 20--30 meter class telescopes \citep[e.g.][]{Gallart2025} would provide crucial benchmarks for assessing the reliability of these estimates.

\subsubsection{Comparison of Sculptor and NGC 6822} \label{subsubsec:acacias_sculptor_ngc6822}

As this work presents the MUSE spectrum analysis of a second dwarf galaxy with \textit{The Cannon} via the ACACIAS program, we are now concerned with the abundance differences of Sculptor and NGC 6822.
Figure~\ref{fig: comparison_ngc6822} compares abundance trends from our preferred models for each element (blue: Full-spectrum; red: Combined) with measurements from \citetalias{Ness2025} for 190 RGB stars in the dwarf irregular galaxy NGC~6822 (pink). Each track represents the running median of each sample, with shaded regions showing the standard deviation around the median. 

This figure illustrates how [Mg/Fe], [Mn/Fe], [Al/Fe], [C/Fe], and [N/Fe] vary with [Fe/H] in these two dwarf galaxies compared to the training set (gray points, Milky Way RGB stars), revealing the chemical diversity of Local Group systems. Each element traces different nucleosynthetic channels and enrichment timescales, providing complementary constraints on the star formation and chemical evolution histories of Sculptor and NGC~6822. 

For [N/Fe], the RMS values in Table~\ref{tab:validation_models} favour the Combined model, as discussed in Section~5.3. However, in the last panel of Figure~\ref{fig: comparison_ngc6822} we show the estimates from both the Combined and Full-spectrum models, as the two show a discrepancy of $\sim 0.28$ dex in their distributions at metallicities below [Fe/H] $< -1.6$. As we mentioned previously, this divergence may reflect the loss of high-signal pixels at lower metallicities in the Combined model, which could reduce the amount of spectral information available to \textit{The Cannon} for estimating [N/Fe] in this regime.

More broadly, these discrepancies between the models at [Fe/H] $< -1.6$ reinforce the need for expanded metal-poor training samples, as both Sculptor and NGC~6822 extend beyond the well-sampled metallicity range of the current training set. Future extensions of ACACIAS to more metal-poor systems will therefore allow for a better representation of the stellar populations found in dwarf galaxies, leading to more accurate and homogeneous abundance estimates.

\section{Conclusions} \label{sec:conclusions}

We present ACACIAS II, the second paper in our program to derive individual element abundances for dwarf spheroidal galaxies using low signal-to-noise and low-resolution spectra \citepalias{Ness2025}. We derive six chemical abundances ([Fe/H], [Mg/Fe], [Mn/Fe], [Al/Fe], [C/Fe], [N/Fe]) for 39 RGB stars in the Sculptor dwarf spheroidal galaxy from $R \sim 1800$, ${\rm SNR} \sim 25$ MUSE spectra. Building upon the initial proof of concept of \citetalias{Ness2025}, we investigate broader questions regarding which modelling approach is most appropriate for these data, and how to effectively scale this method to a wider range of stellar parameters. Using the data-driven code \textit{The Cannon} \citep{Ness2015}, we trained and compared four model variants based on different wavelength selection strategies: (1) the Full-spectrum model (no wavelength restrictions, analogous to \citetalias{Ness2025}), (2) the Absorption model (restricted to known strong absorption lines of each element), (3) the Telluric model (down-weighting regions of telluric contamination), and (4) the Combined model (incorporating both absorption feature selection and telluric masking). Our main findings are as follows:

\begin{itemize}[leftmargin=2em,labelwidth=2em]
    \item We demonstrate that extending the training set to higher surface gravities ($\log g < 4.0$ vs. $\log g < 2.5$ in \citetalias{Ness2025}) and enhanced metal-poor coverage ensure that Sculptor RGB stars fall within the training set parameter space. This expansion enabled \textit{The Cannon} to interpolate rather than extrapolate when predicting stellar labels, as evidenced in Figures~\ref{fig: logg_teff} and \ref{fig: C_logg}. The optimized training set increased the fraction of metal-poor stars ([Fe/H] $< -1.0$ dex) from 5\% to 11\%, providing better representation of dwarf galaxy stellar populations.
    
    \item For estimating [Mg/Fe], [Mn/Fe], [C/Fe], and [N/Fe] abundances, we recommend the Combined model, which achieves precision improvements of 27\%, 18\%, 16\%, and 19\% over the Full-spectrum model, respectively, as evaluated on the training set at a median SNR $\sim200$. These recommendations are broadly consistent with those obtained from the SNR-degraded cross-validation presented in Appendix~\ref{app: degraded_spectra}, which reflects the typical SNR of our MUSE spectra. Table~\ref{tab:validation_models} summarizes the performance metrics for all models.
    
    \item For [Al/Fe], $T_{\rm eff}$, $\log g$, and [Fe/H], we recommend the Telluric model, which achieves 20\% better precision for aluminium compared to the Combined model, evaluated over the training set. The aluminium case demonstrates that restricting to designated Al features excludes correlated information (likely from the Na~\textsc{i} doublet), highlighting that data-driven models can exploit correlations beyond the designated spectral features of each element. However, the resulting label should be interpreted as $\ell_\text{[Al/Fe]}$ rather than a direct [Al/Fe] measurement.

    \item Comparing the Full-spectrum and Combined model predictions for Sculptor reveals generally good agreement with literature measurements for most elements (Figure~\ref{fig: Result: x/fe vs fe/h}). For [Mg/Fe], [Mn/Fe], and [C/Fe], both models recover similar trends, but carbon abundances show systematic offsets from literature samples, reflecting differences in the evolutionary state of our stars. Nitrogen shows large scatter in both our predictions and literature, likely reflecting genuine intrinsic variations or SNR-related uncertainties. Aluminium predictions differ between models, with the Combined model producing higher [Al/Fe] than the Full-spectrum model and literature, confirming that the designated Al features alone are insufficient.

    \item Figure~\ref{fig: comparison_ngc6822} compares abundance patterns between Sculptor (this work) and NGC~6822 (\citetalias{Ness2025}). Discrepancies between the Full-spectrum and Combined models are largest for metal-poor stars, as expected given their limited representation in the training set despite our optimization efforts. Future work will require further augmentation of metal-poor training samples to achieve comparable precision across the full metallicity range of Local Group dwarfs.

\end{itemize}

Looking forward, we plan to extend the ACACIAS framework to additional dwarf galaxies with existing MUSE observations, including Sagittarius I, Leo~T, IC~1613, Pegasus, Tucana, WLM, Sextans~A, and Sextans~B. This will enable a homogeneous abundance census of Local Group dwarf galaxies and Milky Way populations on a consistent scale, allowing studies of chemical enrichment diversity across environments, masses, and star formation histories. By establishing consistent abundance measurements across disrupted and surviving dwarf galaxies, such efforts will provide crucial constraints on the hierarchical assembly of the Milky Way and the nucleosynthetic pathways that built the Galaxy's stellar halo.

\section*{Acknowledgments}

We acknowledge the traditional owners of the land on which the ANU stands, the Ngunnawal and Ngambri people. We pay our respects to elders past, and present and are proud to continue their tradition of surveying the night sky and its mysteries to better understand our Universe.

AC and SB acknowledge support from the Australian Research Council under grant number DE240100150 and the anonymous reviewer for the valuable comments and suggestions.

ChatGPT (OpenAI) was used to suggest language improvements within the manuscript. All scientific content, analysis, interpretations, and references were verified by the authors.

\section*{Data Availability}

All code to reproduce the analysis and figures can be publicly accessed via \url{https://github.com/AnellCornejoCardenas/MuseCannon\_Sculptor}.

\bibliographystyle{mnras}
\bibliography{bib}

\begin{appendix}

\section{Coefficients of different models}\label{app: theta amplitude}

\begin{figure}
    \centering
    \includegraphics[width=\linewidth]{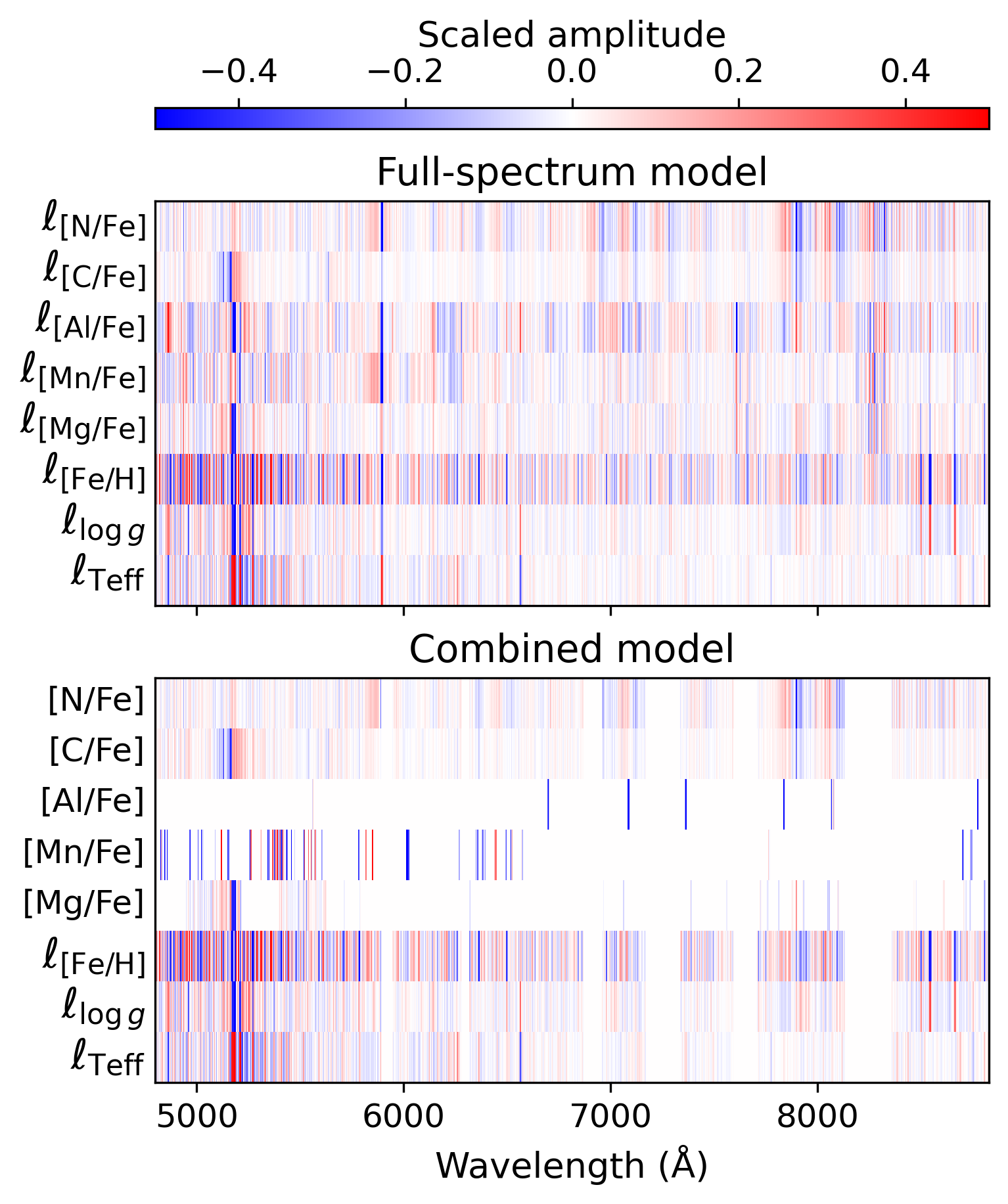}
    \caption{Summary of selected first-order coefficients of the Full-spectrum (top panel) and Combined (bottom panel) models used to infer stellar labels for the Sculptor spectra. Each row is scaled to its absolute maximum value for visualisation purposes. Quantities marked with $\ell$ denote inferred labels.}
    
    \label{fig: theta_amplitudes}
\end{figure}

With \textit{The Cannon}, we infer Sculptor stellar labels from the spectra using four models: Full-spectrum, Absorption, Telluric, and Combined. Each model restricts the wavelength range differently, as explained in the main text, resulting in different sets of first-order $\Theta_\lambda$ coefficients. Figure~\ref{fig: theta_amplitudes} displays the $\Theta_\lambda$ coefficients for the Full-spectrum (upper panel) and Combined (lower panel) models for each training label ($\ell$). 

Following the approach in \citetalias{Ness2025} (their Figure 8), the coefficient amplitudes are normalized by their absolute maximum value across the wavelength region to improve visualization. With this, red regions indicate positive peaks, while blue tones correspond to negative peaks, highlighting the wavelength intervals where the model is most sensitive to variations in each label. White areas mark portions of the spectrum where the model shows little or no sensitivity, indicating wavelengths that contribute minimally to the label inference. Quantities marked with $\ell$ denote inferred labels ($T_{\rm eff}$, $\log g$, and [Fe/H]), derived from correlations across the spectral information used by each model. Element abundances in the Combined model, instead, are constrained by wavelength intervals containing their corresponding absorption features, and therefore are not denoted with $\ell$, as they are more directly linked to the spectral signatures of each element.

The wavelength restrictions imposed on the Combined model are clearly visible in this figure. For chemical abundances ([Mg/Fe], [Mn/Fe], [Al/Fe], [C/Fe], and [N/Fe]), the Combined model extracts information only from wavelengths where known absorption lines of each element are located. For example, aluminium shows sensitivity only in few wavelength regions, reflecting its limited spectral features, while magnesium displays sensitivity across multiple regions due to its stronger absorption signatures. 

In contrast, the Full-spectrum model exhibits distributed sensitivity across the entire spectrum, indicating that label information is extracted from any spectral region without explicit wavelength restrictions.

\section{Cross-validation at low SNR}\label{app: degraded_spectra}

Table \ref{tab:degraded_validation_models} presents the results of an additional cross-validation test designed to assess model performance at the typical SNR of our Sculptor MUSE spectra. While the cross-validation in Table \ref{tab:validation_models} was performed using the original LAMOST spectra (median SNR $\sim 200$), the MUSE spectra of RGB stars in Sculptor have a median SNR $\sim 25$, which may affect the relative performance of the models.

To simulate this regime, we kept the training spectra at their original SNR and degraded only the validation spectra by adding Gaussian noise. Specifically, for each validation spectrum with flux $f$ and inverse variance $\sigma_{\rm current}^{-2}$, we drew additional noise from a Gaussian distribution with standard deviation $\sigma_{\rm add} = \sqrt{\sigma_{\rm target}^2 - \sigma_{\rm current}^2}$, where $\sigma_{\rm target} = f / {\rm SNR}_{\rm target}$ is the target uncertainty corresponding to the median SNR of our MUSE spectra, with ${\rm SNR}_{\rm target} = 25$. We chose a fixed target SNR rather than sampling from the Sculptor SNR distribution in order to keep the test general and not tied exclusively to the properties of our Sculptor sample, making the results applicable to other datasets with similar SNR regimes.

\begin{table*}
\centering
\caption{Bias and RMS scatter around the 1:1 line for each model's predictions relative to APOGEE reference values, evaluated on the SNR-degraded test set.}
\begin{tabular}{lcccccc}
\hline \hline
Label & Full-spectrum model & Absorption model & Telluric model & Combined model \\ 
\hline
Teff & \boldmath{$~~2 \pm 76$} & $~-2 \pm 80$ & $~~0 \pm 77$ & $~-2 \pm 79$ \\ 
logg & $-0.00 \pm 0.19$ & $-0.01 \pm 0.21$ & \boldmath{$-0.00 \pm 0.19$} & $-0.01 \pm 0.20$ \\ 
Fe & $0.001 \pm 0.083$ & $-0.002 \pm 0.088$ & \boldmath{$-0.001 \pm 0.083$} & $-0.004 \pm 0.085$ \\ 
Mg & $0.000 \pm 0.079$ & $0.006 \pm 0.079$ & \boldmath{$-0.000 \pm 0.076$} & $0.007 \pm 0.079$ \\ 
Mn & $0.00 \pm 0.13$ & $0.00 \pm 0.12$ & $-0.00 \pm 0.13$ & \boldmath{$0.00 \pm 0.12$} \\
Al & \boldmath{$0.01 \pm 0.13$} & $0.02 \pm 0.27$ & $0.01 \pm 0.14$ & $0.02 \pm 0.27$ \\
C & $-0.006 \pm 0.097$ & $-0.005 \pm 0.092$ & $-0.006 \pm 0.095$ & \boldmath{$-0.006 \pm 0.092$} \\
N & $-0.00 \pm 0.15$ & $0.01 \pm 0.14$ & $-0.00 \pm 0.15$ & \boldmath{$0.00 \pm 0.13$} \\ 
\hline \hline
\end{tabular}
\label{tab:degraded_validation_models}
\end{table*}

Overall, the results are broadly consistent with the recommendations presented in the main text, where the Telluric and Combined models achieve the best performance for estimating the labels considered in this work from spectra with SNR $\sim 25$.

For the stellar parameters ($T_{\rm eff}$, $\log g$, and [Fe/H]), the Telluric model remains among the best-performing models in Table~\ref{tab:degraded_validation_models}, consistent with the main text. However, at this lower SNR regime, the Full-spectrum model shows a more comparable performance, and for $T_{\rm eff}$ specifically, it achieves an RMS $\sim 1\%$ lower than the Telluric model. This suggests that at lower SNR, \textsl{The Cannon} continues to benefit from information across the full spectrum when estimating these labels, which have a broad spectral dependence.

For the chemical abundances, the main text recommends the Combined model for all labels except [Al/Fe]. The results in Table~\ref{tab:degraded_validation_models} are consistent with this recommendation: the Combined model maintains good performance for [Mn/Fe], [C/Fe], and [N/Fe]. For [Mg/Fe], the Telluric model achieves a slightly lower RMS than the Combined model at this SNR, suggesting a marginally better performance on low-SNR spectra. However, this difference amounts to only $\sim 4\%$. For [Al/Fe], the models that draw information from the full spectrum again outperform those restricted to the designated Al absorption lines (the Absorption and Combined models), consistent with the conclusion in the main text that \textit{The Cannon} exploits the Na~\textsc{i} doublet to estimate aluminium abundances.

Taken together, these results support the conclusion drawn in the main text that caution should be exercised when interpreting abundances derived from low-SNR spectra, particularly for [Al/Fe], and reinforce the value of high-resolution follow-up observations as benchmarks for validating these estimates. It should be noted that this analysis was performed at a fixed target SNR $\sim 25$, and additional caution is warranted at lower SNR regimes. This is especially relevant for the Combined model, which restricts \textsl{The Cannon} to a small fraction of the spectral pixels associated with each element, making it essential that the flux in those pixels carries sufficient signal above the noise level to reliably constrain each label.

\section{Sculptor individual abundances}\label{app: sculptor individual estimations}

\begin{figure}
    \centering
    \includegraphics[width=1\linewidth]{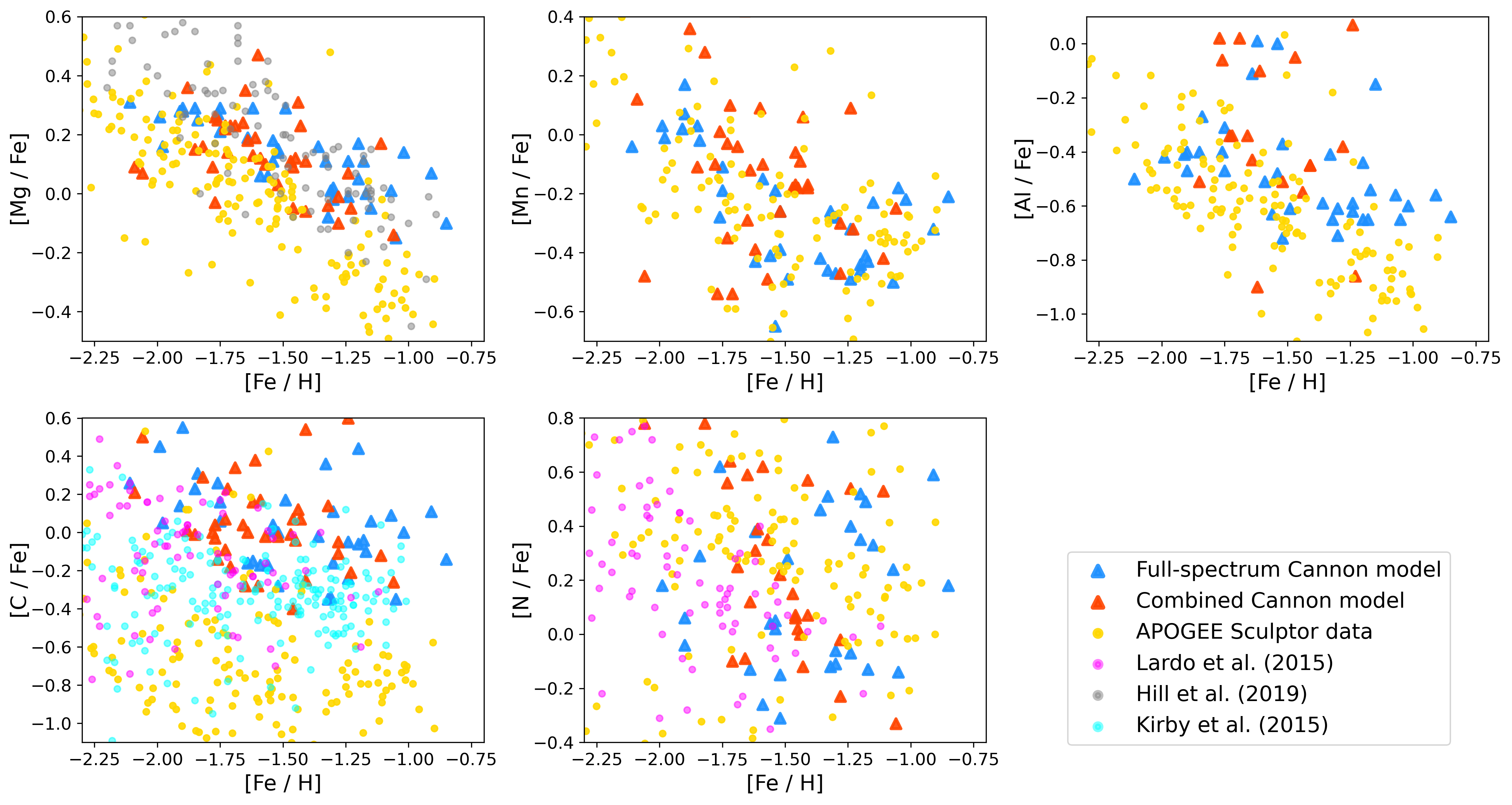}
    \caption{Scatter distribution of the diagrams of [X/Fe] as a function of [Fe/H] for the 39 stars in Sculptor dwarf galaxy. The tracks in Figure \ref{fig: Result: x/fe vs fe/h} correspond to the medians in bins of [Fe/H] from these distributions.} 
    \label{fig: sculptor_individual_scatter}
\end{figure}

To ensure transparency in our measurements and facilitate comparison with future studies, Figure~\ref{fig: sculptor_individual_scatter} presents the individual [X/Fe] abundance estimates for all 39 Sculptor stars derived from the Combined (red triangles) and Full-spectrum (blue triangles) models, without binning or averaging. These measurements complement the binned analysis presented in Figure~\ref{fig: Result: x/fe vs fe/h}, which shows median abundances and percentile ranges. 

\section{Examples of modelled versus observed spectra}\label{app: modelled_spectra}

\begin{figure}
    \centering
    \includegraphics[width=1\linewidth]{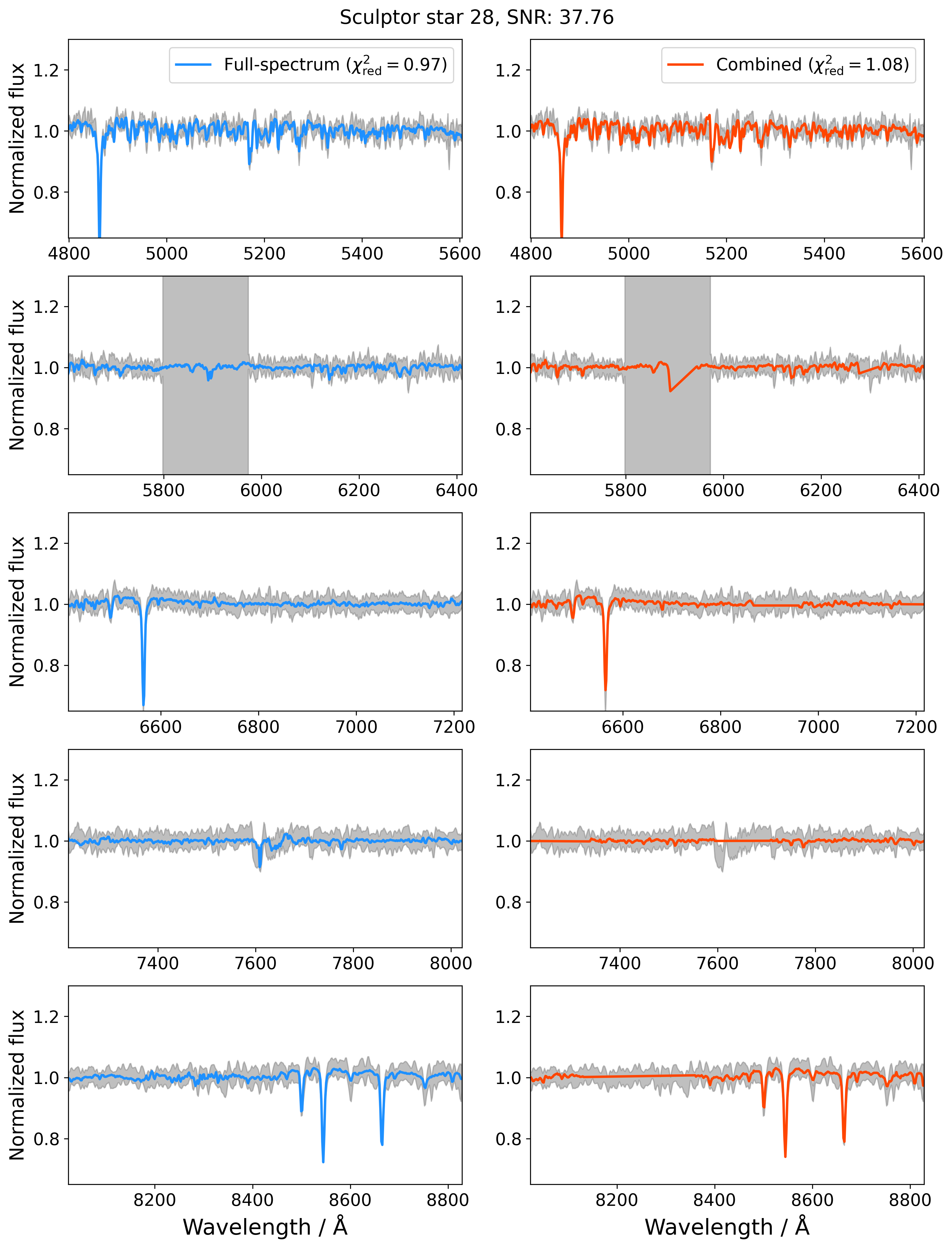}
    \caption{ Comparison between the spectra generated by the Full-spectrum and Combined models (blue and red, respectively) with the observed MUSE spectrum (gray) for a Sculptor star with SNR=37.76. Each row covers a different wavelength range.
    } 
    \label{fig: spectra_comparison_highsnr}
\end{figure}

\begin{figure}
    \centering
    \includegraphics[width=1\linewidth]{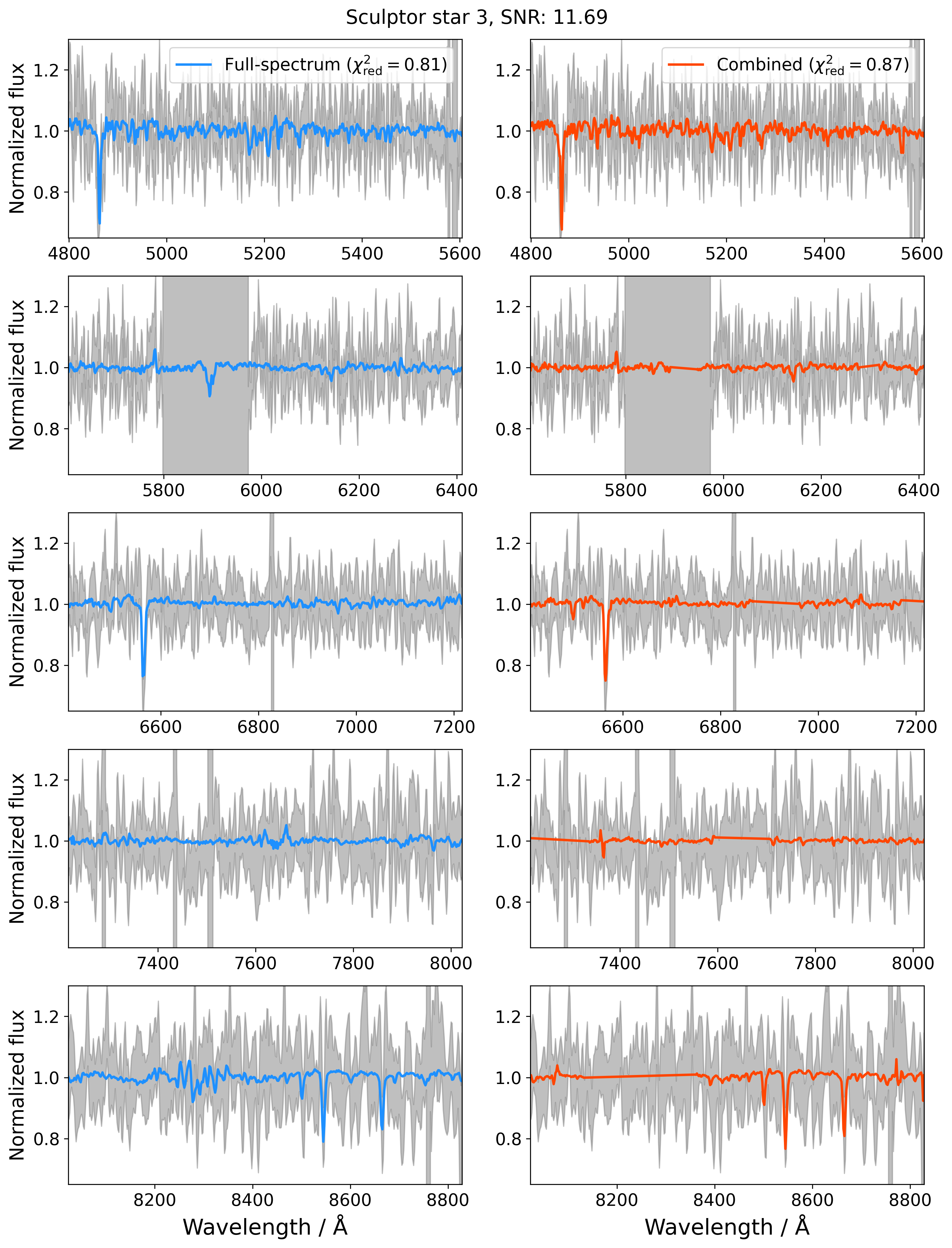}
    \caption{Same as Fig.~\ref{fig: spectra_comparison_highsnr}, but considering a Sculptor star with SNR=11.69.} 
    \label{fig: spectra_comparison_lowsnr}
\end{figure}

Figures~\ref{fig: spectra_comparison_highsnr} and \ref{fig: spectra_comparison_lowsnr} show representative spectral comparisons for two stars in our Sculptor sample with different SNRs. In each figure, the left panels show the spectrum generated by the Full-spectrum model (blue) and the right panels show the spectrum generated by the Combined model (red). The observed MUSE spectrum is shown as a gray shaded region spanning $\pm 1\sigma$ around the observed flux, and the reduced $\chi^2$ value is reported for each modelled spectra. The reduced $\chi^2$ values shown for these individual examples are close to unity and illustrate the good agreement between the observed and modelled spectra.

\end{appendix}

\bsp	
\label{lastpage}
\end{document}